\RequirePackage{etoolbox}
\csdef{input@path}{%
 {sty/}
 {img/}
}%
\csgdef{bibdir}{bib/}

\documentclass[ba]{imsart}
\pubyear{0000}
\volume{00}
\issue{0}
\doi{0000}
\firstpage{1}
\lastpage{1}

\usepackage{amsthm}
\usepackage{amsmath}
\usepackage{natbib}
\usepackage[colorlinks,citecolor=blue,urlcolor=blue,filecolor=blue,backref=page]{hyperref}
\usepackage{graphicx}

\startlocaldefs
\numberwithin{equation}{section}
\theoremstyle{plain}

\endlocaldefs

\begin{document}

\begin{frontmatter}
\title{Bayesian Mixture Models With Focused Clustering for Mixed Ordinal and Nominal Data\thanksref{T1}}
\runtitle{Focused Clustering for Mixed Data}

\begin{aug}
\author{\fnms{Maria} \snm{DeYoreo}\thanksref{addr1,t1,t2}\ead[label=e1]{maria.deyoreo@stat.duke.edu}},
\author{\fnms{Jerome P.} \snm{Reiter}\thanksref{addr1,t1,t2}\ead[label=e2]{jerry@stat.duke.edu}}
\and
\author{\fnms{D. Sunshine} \snm{Hillygus}\thanksref{t1,addr2}
\ead[label=e3]{hillygus@duke.edu}
}

\runauthor{M. DeYoreo et al.}

\address[addr1]{Department of Statistical Science, Duke University, Durham, NC, USA
    \printead{e1} 
    \printead{e2}
}

\address[addr2]{Department of Political Science, Duke University, Durham, NC, USA
    \printead{e3}
}

\thankstext{t1}{M. DeYoreo is postdoctoral researcher, J. P. Reiter is Professor of Statistical Science, and D. S. Hillygus is Professor of Political Science, Duke University.}
\thankstext{t2}{This research was supported in part by \textit{The National Science Foundation} under 
award SES-11-31897.}

\end{aug}

\begin{abstract}
In some contexts, mixture models can fit certain variables well at the expense of others
in ways beyond the analyst's control.  For example, when the data include some
variables with non-trivial amounts of missing values, the mixture
model may fit the marginal distributions of the nearly and fully complete variables at the
expense of the variables with high fractions of missing data.  
Motivated by this setting, we present a mixture model for mixed ordinal and
nominal data that splits variables into two
groups, focus variables and remainder variables.  
The model allows the analyst to specify a rich sub-model for the focus variables and a simpler
sub-model for remainder variables, yet still capture
associations among the variables. 
Using simulations, we illustrate advantages and
limitations of 
focused clustering compared to mixture models that do not distinguish
variables. We apply the model to handle missing values in an analysis
of the 2012 American National Election Study, estimating relationships among voting behavior, ideology, and
political party affiliation.   
\end{abstract}



\end{frontmatter}

\section{Introduction}
\label{sec:intro} 

Many government and social science surveys include a mix of ordered
and nominal categorical variables.  Typically, these surveys suffer
from missing values due to item nonresponse.  To deal with the
complications that result, common strategies
include analyzing only the complete cases, which leads to inefficient and potentially biased
inferences \citep{little:rubin:2002},  using multiple imputation  in
advance of likelihood-based or survey-weighted 
inference on the completed datasets \citep{rubin}, and using Bayesian models
that integrate over the missing data.  For the latter two approaches,
mixture models 
are particularly effective and computationally convenient engines for
imputation and inference
\citep{si,muller:mitra,manrique,deyoreo:binary}.

While 
mixture models have the potential to capture complex
  dependencies, in practice they may fit the distribution of certain sets of
variables at the expense of other sets \citep{hannah,banerjee,wadedunson,murray}.  
For example, when the data comprise many nominal variables and a small number
of ordinal variables, the model might seek clusters that estimate the
distribution of the nominal variables as best as possible, but in the
process sacrifice the fit of the ordinal variables \citep{murray}.  Additionally,
standard mixture models often capture dependence among variables only through
clustering. This may demand a large number of mixture
components, possibly more than the data can estimate reliably. 
Similar  problems are encountered in joint modeling for
regression when the covariates are high-dimensional compared to the
response variables. The creation of a large number of 
mixture components in order to fit the marginal distribution of
the covariates accurately can lead to poor predictive inference \citep{hannah,wadedunson,Petralia}.


These types of practical problems can be compounded when the data include some
variables with non-trivial amounts of missing values. With  modest
sample sizes,  
mixture  models may fit the marginal distributions
of  the nearly and fully complete variables at the
expense of the variables with high fractions of missing data.
When using the model for multiple imputation, this is exactly the
opposite of what we want: the
quality of the imputation model is particularly important for variables
missing at high rates and less important for variables
missing at low rates or that are completely observed.  Related, suppose that in
a database with $p$ variables, an analyst seeks to estimate the joint
distribution of a 
particular subset of $q < p$ variables as accurately as possible. When
$q$ is small compared to $p$,  fitting a model to all $p$ variables
can waste fitting power on the $p-q$ less important
variables. Nonetheless, the analyst may not want to completely throw away the
information in the $p-q$ variables, which can be useful for predicting
missing values among the $q$ variables of interest \citep{rubin1996}.

In this article, we present an approach for joint modeling of mixed
ordinal and nominal data intended to address these issues.  The basic idea is to split variables into two
groups, {\em focus variables} and {\em remainder variables}. For example, in missing data contexts, the focus
variables might include key variables with high rates of missing
values, and the remainder variables might include variables without
much missing data.  The partitioning allows us to fit a rich mixture sub-model
for the focus variables and a relatively simple mixture sub-model for the marginal distribution 
of the remainder variables, thereby focusing fitting power where it is most desired.  
We induce dependence between 
the focus and remainder variables in two ways.  First, we use a multivariate ordered probit specification \citep{albert,chibgreen} 
for the ordinal focus variables and allow the means to depend on functions of the remainder variables.  Second, we 
use a tensor factorization (TF) prior \citep{banerjee} to make cluster assignments in both sub-models dependent.  A similar 
strategy is used by \citet{murray} in a mixture model for nominal and continuous data without focused clustering. 
We call the integrated model a {\em mixture model with focused clustering}, abbreviated as MM-FC.

The remainder of this article is organized as follows.  
In Section \ref{sec:methodology}, we begin by motivating the benefits of using
mixture models for modeling and multiple imputation with mixed ordinal and
nominal categorical data.
We then describe the MM-FC and its properties. In
Section \ref{sec:simulations}, we present results of simulation studies in
which we assess the performance of a MM-FC that separates variables into groups based on
degree of missingness. We consider different scenarios related to rate of missingness, sample size, and
number of focus variables. In Section \ref{sec:ANES}, we use the MM-FC
to create multiply-imputed datasets from the 2012 American National
Election Study (ANES), and make inferences on relationships
among voting behavior, ideology, and political party affiliation.   In
Section \ref{sec:discussion}, we conclude with a discussion of future research
directions.


\section{Motivation for and Specification of the MM-FC}\label{sec:methodology}

\subsection{Motivation}\label{sec:exploration}

When modeling the joint distribution of categorical data, one standard approach
is to estimate a log-linear model \citep{bishop}.
This effectively treats any ordinal variable as nominal, which
sacrifices information in the ordering.  Perhaps more importantly, 
with many variables the space of possible log-linear models is
enormous, and it is difficult to determine which interaction effects
to include in the linear predictor \citep{vermunt,si}.  Simple main effects or two-way interactions models
often are inadequate to describe relationships among survey
variables, especially in social science data. For
example, in the ANES, a log-linear model with all
two-way interaction terms is insufficient for describing relationships among party,
vote intent, and ideology ($\chi^2$--test p--value $< .01$), which are included in many analyses in political science. 

Another approach is to form a joint model as a product of conditional
distributions, e.g., $f(x, y, z) = f(x)f(y \mid x) f(z \mid x, y)$, as
suggested by \citet{lipsitz} and \citet{ibrahim:1999}. As the
  number of variables in the conditioning set increases, specifying
  the conditional model becomes increasingly challenging. It can be difficult to
select the interactions that should enter any particular model,
  particularly when the data have few complete cases that one can use to
  search for interactive relationships.    
Further, when using multinomial probit regressions as 
conditional models, it may not be realistic to assume that the ordinal
outcomes have underlying latent continuous variables
that are normally distributed \citep{boes}.  
To illustrate, suppose that interest in the ANES data centers
on how congressional approval $Y$ (levels $1$ to $4$) varies with
ideology $X$. A standard probit model implies that $\text{Pr}(Y=1\mid
X)$ has the opposite type of monotonicity from $\text{Pr}(Y=4\mid X)$
as a function of $X$. However, ANES data suggest that both trends are
unimodal, favoring moderate values. As discussed by \citet{kottas},
the multivariate probit model is inappropriate for data that does not
concentrate most of its data in central cells. The ANES data contain
many ordinal variables that refer to opinions on various topics, and
people are often more likely to fall into one of the extreme
categories indicating strong feelings than the moderate categories
indicating lack of feelings or opinions (e.g., opinion on President Obama, Congress, health care). 

\subsection{The MM-FC Model}\label{sec:mm-fc}

The strategy of focused clustering can be applied with any type of mixture kernels for the sub-models.
Here, we present a model motivated by missing data contexts, where we
seek a rich distribution for multiple imputation of variables with
high amounts of missingness and are willing to accept a simpler
specification for the marginal distribution of variables with few or
no missing values. We use a hierarchically coupled mixture model with local dependence \citep{murray} for the 
ordinal and nominal focus variables, and a computationally convenient
finite mixture of independent multinomial kernels \citep{dunsonxing} for the remainder variables.  
We note that one also could account for  ordinality in the remainder variables;
details on this alternative formulation are in the supplementary material.


In this section, we present the model for data that are fully observed---values that are missing at random (MAR) can be handled in the MCMC sampler---saving explanations of various model choices and properties for Section \ref{properties}. 
We suppose that the data comprise $n$ individuals measured on a total of $p$ ordered categorical and nominal variables. 
We split the $p$ variables into $p_A$ focus variables referenced by set ${A}$ and $p_B$ remainder variables referenced by set ${B}$. 

Within $A$, we suppose that there are $p_{Ac}$ ordered categorical variables and $p_A - p_{Ac}$ nominal variables. 
Let $Y_{ij}^{(A)} \in \{1, \dots, L_j\}$ be the value of ordered categorical variable $j\in \{1,\dots,p_{Ac}\}$ for individual $i$. 
Following the usual multivariate probit specification,  
for $i=1,\dots, n$ and $j = 1, \dots, p_{Ac}$, let $Z_{ij}^{(A)}$ be a latent continuous variable corresponding to $Y_{ij}^{(A)}$. 
Let $\mathbf{Y}_i^{(A)}=(Y_{i1}^{(A)},\dots,Y_{ip_{Ac}}^{(A)})$,
and let 
$\mathbf{Z}_i^{(A)}=(Z_{i1}^{(A)},\dots,Z_{ip_{Ac}}^{(A)})$.
For $i=1, \dots, n$ and $j = p_{Ac}+1, \dots, p_A$, let $X_{ij}^{(A)} \in \{1, \dots, L_j\}$ be
the value of nominal focus variable $j$ for individual $i$, and let 
$\mathbf{X}^{(A)}_i=(X_{ip_{Ac}+1}^{(A)},\dots,X_{ip_{A}}^{(A)})$. 
Treating all $p_B$ variables within $B$ as nominal, for $i=1, \dots, n$ and $j=p_A+1,\dots,p$, let $X_{ij}^{(B)} \in \{1, \dots, L_j\}$ be the value of  remainder variable $j$ for individual $i$. Let 
$\mathbf{X}^{(B)}_i=(X_{ip_{A}+1}^{(B)},\dots,X_{ip}^{(B)})$.  
Thus, the data for individual $i$ are $(\mathbf{Y}_i^{(A)},\mathbf{X}_i^{(A)},\mathbf{X}_i^{(B)})$, where $\mathbf{Y}_i^{(A)}$ is indexed by $j=1,\dots,p_{Ac}$, $\mathbf{X}_i^{(A)}$ is indexed by $p_{Ac}+1,\dots,p_A$, and $\mathbf{X}_i^{(B)}$ is indexed by $p_A+1,\dots,p$.

To enable focused clustering, we introduce distinct allocation variables for variables in ${A}$ and in ${B}$.
Furthermore, following \citet{murray}, we introduce separate mixture component indices for each data type within
${A}$.  For $i=1, \dots, n$, let $H_i^{(ZA)}$ be
the $i$th individual's label of the mixture component for the ordered categorical  focus
variables (via the latent continuous variables); let $H_i^{(XA)}$ be
the label of the mixture component for the nominal focus variables;
and, let  $H_i^{(B)}$ be the label of the mixture component for the 
remainder variables. We assume  that $(H_i^{(ZA)}, H_i^{(XA)}, H_i^{(B)})$  
arise from discrete distributions supported on $\{1,\dots,N^{(ZA)}\}$,
$\{1,\dots,N^{(XA)}\}$, and $\{1,\dots,N^{(B)}\}$, respectively. We discuss how to choose these truncation levels in the supplementary material.

The data model combines multivariate normal kernels  
\citep[e.g., as in][]{bohning,elliott,kim,kim:JASA} with multinomial
kernels \citep[e.g., as in][]{dunsonxing, si}.  Applying an ordinal probit specification to
handle the ordered categorical variables in $A$, and letting $\mathbf{X}_i=(\mathbf{X}^{(A)}_i,\mathbf{X}^{(B)}_i)$, we have  
\begin{eqnarray} \label{eqn:data_model}
(\mathbf{Z}_i^{(A)}\mid \mathbf{X}_i,H_i^{(ZA)}=r, \boldsymbol{\beta}_r, \boldsymbol{\Sigma}_r)\stackrel{ind}{\sim} \mathrm{N}_{p_{Ac}}(\mathbf{Z}_i^{(A)};\boldsymbol{D}(\mathbf{X}_i)\boldsymbol{\beta}_{r},\boldsymbol{\Sigma}_{r}), \, i=1,\dots, n \\
(X_{ij}^{(A)}\mid
H_i^{(XA)}=l, \boldsymbol{\psi}^{(j)}_l)\stackrel{ind}{\sim}
\mathrm{categ}(\boldsymbol{\psi}_{l}^{(j)}), \quad  i=1,\dots,n, \,
j=p_{Ac}+1,\dots, p_{A}\label{model:eq2} \\
(X_{ij}^{(B)}\mid
H_i^{(B)}=s, \boldsymbol{\phi}^{(j)}_s)\stackrel{ind}{\sim}
\mathrm{categ}(\boldsymbol{\phi}_{s}^{(j)}), \quad  i=1,\dots,n, \,
j=p_{A}+1, \dots, p. \label{model:eq4}\end{eqnarray}
Here, $\boldsymbol{D}(\mathbf{X}_i)$ is a design vector of length $d$, and
$\boldsymbol{\beta}_r$ is a $d\times p_{Ac}$ matrix of regression coefficients. We discuss specification of  $\boldsymbol{D}(\mathbf{X}_i)$ in Section \ref{properties}.
The notation $\mathrm{categ}(\cdot)$ denotes a categorical distribution, i.e., if $X\sim \mathrm{categ}(p_1,\dots,p_k)$, then $\mathrm{Pr}(X=i)=p_i$, for $i=1,\dots,k$.

We let $Y_{ij}^{(A)}=k$ if and only if
$\gamma_{j,k-1}^{(A)}<Z_{ij}^{(A)}\leq \gamma_{j,k}^{(A)}$, for
$j=1,\dots,p_{Ac}$ and $k=1,\dots,L_j$. 
The cut-off points
$(\gamma_{j,1}^{(A)},\dots,\gamma_{j,L_{j}-1}^{(A)})$, where 
$-\infty =\gamma_{j,0}^{(A)}<\gamma_{j,1}^{(A)}<\cdot \cdot \cdot
<\gamma_{j,L_{j}-1}^{(A)}<\gamma_{j,L_j}^{(A)}=\infty$, can be fixed
to arbitrary increasing values, which we recommend to be centered at
zero and equally spaced \citep{kottas,deyoreokottas,bao}. This is an
attractive property of the mixture model, as the cut-off points are computationally
difficult to estimate when treated as random.

While we focus on settings with discrete variables only, 
one can include continuous focus variables in the multivariate normal kernel for $\mathbf{Z}^{(A)}$ and not treat them as latent. Continuous remainder variables can be incorporated via independent normal kernels in the model for $B$. See \citet{canale} for a related model that deals with
  mixed continuous, count, and ordinal data, modeled jointly through a
  multivariate normal kernel. As with the ordinal variables in our
  model, the discrete realizations are obtained through latent
  continuous random variables by partitioning of the real line.

We model $(H_i^{(ZA)}, H_i^{(XA)}, H_i^{(B)})$ as conditionally independent given another
set of components, $H_i \in \{1, \dots, N\}$. For $h=1,\dots,N$,
we have $\mathrm{Pr}(H_i^{(ZA)}=r\mid H_i=h)=\pi_{rh}^{(ZA)}$ where
$r=1,\dots,N^{(ZA)}$, $\mathrm{Pr}(H_i^{(XA)}=l\mid
H_i=h)=\pi_{lh}^{(XA)}$ where $l=1,\dots,N^{(XA)}$, and
$\mathrm{Pr}(H_i^{(B)}=s\mid H_i=h)=\pi_{sh}^{(B)}$ where
$s=1,\dots,N^{(B)}$. 
We assume that $\mathrm{Pr}(H_i=h)=\pi_{h}$ for all $h$.
All these 
probabilities are determined through stick-breaking of latent beta
distributed random variables, 
 defined as
\begin{eqnarray}\label{eqn:prior_weights}
\pi_{rh}^{(ZA)}=V_{rh}^{(ZA)}\prod_{k=1}^{r-1}(1-V_{kh}^{(ZA)}), \quad r=1,\dots,N^{(ZA)}, \, h=1,\dots,N \label{eq5}  \\
\pi_{lh}^{(XA)}=V_{lh}^{(XA)}\prod_{k=1}^{l-1}(1-V_{kh}^{(XA)}), \quad l=1,\dots,N^{(XA)}, \, h=1,\dots,N \label{eq6}\\
\pi_{sh}^{(B)}=V_{sh}^{(B)}\prod_{k=1}^{s-1}(1-V_{kh}^{(B)}), \quad
s=1,\dots,N^{(B)}, \, h=1,\dots,N\label{eq7}\\
\pi_{h}=V_{h}\prod_{k=1}^{h-1}(1-V_{k}), \quad  h=1,\dots, N \label{eq8}\\
V^{(ZA)}_{rh}\mid \alpha^{(ZA)}\stackrel{iid}{\sim}\mathrm{beta}(1,\alpha^{(ZA)}),\quad r=1,\dots,N^{(ZA)}-1, \, h=1,\dots,N \label{eq9}\\
V^{(XA)}_{lh}\mid \alpha^{(XA)}\stackrel{iid}{\sim}\mathrm{beta}(1,\alpha^{(XA)}),\quad l=1,\dots,N^{(XA)}-1, \, h=1,\dots,N \label{eq10}\\
V^{(B)}_{sh}\mid \alpha^{(B)}\stackrel{iid}{\sim}\mathrm{beta}(1,\alpha^{(B)}),\quad s=1,\dots,N^{(B)}-1, \, h=1,\dots,N \label{eq11} \\
V_{h}\mid \alpha\stackrel{iid}{\sim}\mathrm{beta}(1,\alpha), \quad h=1,\dots,N-1. \label{eq12}
\end{eqnarray}
As a consequence of the finite truncation approximation to the DP prior that induces the weights, each $V^{(ZA)}_{N^{(ZA)}h}=1$, for $h=1,\dots,N$.  This makes each
vector $\{\pi^{(ZA)}_{1h},\dots,\pi^{(ZA)}_{N^{(ZA)}h}\}$ sum to
$1$. This also holds for the variables having superscripts $(XA)$ and $(B)$. Additionally, $V_N=1$.

The MM-FC includes an extension of the model of \citet{murray} that accommodates ordinal data
as a special case. We obtain the extension by removing (\ref{model:eq4}), 
placing all latent continuous variables in $\mathbf{Z}^{(A)}$ and all nominal variables in $\mathbf{X}^{(A)}$.
We refer to this model as MM-Mix. As there are no ${B}$ variables in this model, lines (\ref{eq7}) and (\ref{eq11}) are also removed.

For the MM-FC, we use conjugate base distributions for all mixing parameters. These are given by
\begin{eqnarray}\label{eqn:prior_model}
\boldsymbol{\beta}_r\mid \boldsymbol{B_0},\boldsymbol{\tau}\stackrel{iid}{\sim}\mathrm{MN}_{d\times p_{Ac}}(\boldsymbol{B_0},\boldsymbol{I}_d,\mathrm{diag}(\tau_1^2,\dots,\tau_{p_{Ac}}^2)), \quad r=1,\dots,N^{(ZA)}\\
\boldsymbol{\Sigma}_r\mid \boldsymbol{S}\stackrel{iid}{\sim}\mathrm{IW}(\nu,\boldsymbol{S}), \quad r=1,\dots,N^{(ZA)}\\
\boldsymbol{\psi}^{(j)}_l\stackrel{ind}{\sim}\mathrm{Dirichlet}(a^{(\psi j)}_1,\dots,a^{(\psi j)}_{L_j}), \quad j=p_{Ac}+1,\dots, p_A, \, l=1,\dots,N^{(XA)}\nonumber\\
\boldsymbol{\phi}^{(j)}_s\stackrel{ind}{\sim}\mathrm{Dirichlet}(a^{(\phi j)}_1,\dots,a^{(\phi j)}_{L_j}), \quad j=p_{A}+1,\dots, p_B, \, s=1,\dots,N^{(B)}
\end{eqnarray}
where $\mathrm{MN}_{d\times p_{Ac}}$ denotes a matrix-normal
distribution of dimension $d$ by $p_{Ac}$. This implies that
  $\mathrm{vec}(\boldsymbol{\beta}_r)\sim
  \mathrm{N}_{dp_{Ac}}(\mathrm{vec}(\boldsymbol{B_0}),
  \mathrm{diag}(\tau_1^2,\dots,\tau_{p_{Ac}}^2)\otimes
  \boldsymbol{I}_d)$, where vec($\boldsymbol{\beta}_r$) denotes the
  vectorization of $\boldsymbol{\beta}_r$, obtained by stacking its
  columns. Hyperprior specification and posterior inference is discussed in the supplementary material. An alternative to the conjugate inverse--Wishart prior base
  distribution for $\boldsymbol{\Sigma}_r$ is 
  to model $\mathbf{Z}^{(A)}$ with a mixture of factor analyzers
  \citep{ghahramani,gorur,mcparland}. This can be particularly useful when
  $\boldsymbol{\Sigma}_r$ is of high dimension, although this is not
  the case in the ANES data that we analyze. We further discuss this
  alternative model specification  in the supplementary material.

\subsection{Modeling Choices and Model Properties}\label{properties}

To motivate some of the modeling choices behind MM-FC, it is
instructive to compare MM-FC with other approaches that
might be considered for mixed ordinal and nominal data.  We begin with
other models that implement focused clustering. 

One could completely disregard the ordinal nature of $\mathbf{Y}^{(A)}$ and use 
mixtures of independent multinomial distributions for both focus and
remainder variables.  However, evidence from \citet{murray} and our
own simulation studies suggest that the sub-model in Section
\ref{sec:mm-fc} for the focus variables  can estimate the joint
distribution of the focus variables more accurately than a finite
mixture of independent multinomials sub-model; see the supplementary
material for corroborating simulations. Hence, we prefer to take the
ordinal nature of $\mathbf{Y}^{(A)}$ into account.

Alternatively, one could use a single shared cluster index for
$\mathbf{Z}^{(A)}$ and $\mathbf{X}^{(A)}$, say $H^{(A)}$, and set
$\mathrm{E}(\mathbf{Z}^{(A)} \mid H^{(A)}=h) = \boldsymbol{\mu}_h$, making
$\mathbf{Z}^{(A)}$ locally independent of $\mathbf{X}$.  This would
force all associations between $\mathbf{Z}^{(A)}$ and $\mathbf{X}^{(A)}$
to be captured by the clustering in $A$, and all associations between
$\mathbf{Z}^{(A)}$ and $\mathbf{X}^{(B)}$ to be captured by the
dependent cluster assignments in the TF prior.  This is a significant
challenge when complicated relationships and distributions are
present.  As a result, this specification can require a large number of mixture
components, and therefore a large sample size, for accurate
estimation. When sample sizes are modest, the resulting inferences can
be degraded; see \citet{banerjee} for discussion of this issue for similar classes of
mixture models.  
Hence, we prefer to use separate but dependent
cluster variables $(H^{(ZA)}, H^{(XA)}, H^{(B)})$, and allow the means
of $\mathbf{Z}^{(A)}$ to depend locally on $\mathbf{X}$.


Of course, one could completely eschew focused clustering and simply use
the sub-model for the focus variables for all of $(\mathbf{Y}^{(A)},
\mathbf{X})$.  In the simulation studies, we compare MM-FC against its 
closest analogue that does not used focused clustering, MM-Mix.


In MM-FC as well as MM-Mix, the role of
$\boldsymbol{D}(\mathbf{X})$ is to help the model capture
dependence between $\mathbf{Y}^{(A)}$ and $\mathbf{X}$, so that
the mixture components, and especially the TF prior, need not do all
the heavy lifting. Naturally, the functional form of 
$\boldsymbol{D}(\mathbf{X})$ that generates the most useful results,
e.g., imputations that come from a close  approximation of the true
joint distribution, is specific to the data at hand.  We recommend
starting with main effects of each variable in $\mathbf{X}$ as a
default specification. The analyst can evaluate the suitability of
$\boldsymbol{D}(\mathbf{X})$ by looking for evidence of missed
interaction effects.  For example, if draws from the posterior predictive
distribution of $\mathbf{Y}^{(A)}$ for some combination of variables
in $\mathbf{X}$ are quite different from the corresponding observed data
distribution, the model may benefit from adding interactions between
those variables.  
We use such checks in Section \ref{sec:ANES} in the analysis of the
ANES data.


We now turn to specific properties of MM-FC.  
Marginalizing over the mixture allocation indicator variables, the joint density for all variables can be expressed as
$f(\mathbf{Z}^{(A)},\mathbf{X}^{(A)},\mathbf{X}^{(B)}) =$  
\begin{equation}\label{eqn:joint_dens}
 \begin{split}
\sum_{h=1}^N \pi_h  & \left(\sum_{r=1}^{N^{(ZA)}}\pi_{rh}^{(ZA)}\mathrm{N}(\mathbf{Z}^{(A)}; \boldsymbol{\beta}_r \boldsymbol{D}(\mathbf{X}),\boldsymbol{\Sigma}_r)\right)\left(\sum_{l=1}^{N^{(XA)}}\pi_{lh}^{(XA)}\prod_{j=p_{Ac}+1}^{p_A}\mathrm{categ}(X_j^{(A)};\boldsymbol{\psi}_l^{(j)})\right) \\
&  \times
\left(\sum_{s=1}^{N^{(B)}}\pi_{sh}^{(B)}\prod_{j=p_A+1}^{p}\mathrm{categ}(X_j^{(B)};\boldsymbol{\phi}_s^{(j)})\right).\end{split}\end{equation} 
This is a mixture with $N$ components, where each component takes the
form of a product of three mixture models, one for each of
$\mathbf{Z}^{(A)}$, $\mathbf{X}^{(A)}$, and
$\mathbf{X}^{(B)}$. 

The sub-model corresponding to 
$f(\mathbf{Z}^{(A)}\mid \mathbf{X}^{(A)},\mathbf{X}^{(B)})$ is a mixture of
multivariate normal linear regressions, with means and weights that are functions of
$\mathbf{X}$.  In particular,  we have $f(\mathbf{Z}^{(A)}\mid
\mathbf{X}^{(A)}=\mathbf{x}^{(A)}, \mathbf{X}^{(B)}=\mathbf{x}^{(B)})=$  
\begin{equation}\label{eqn:weightsum}\sum_{r=1}^{N^{(ZA)}}\frac{w_r(\mathbf{x}^{(A)},\mathbf{x}^{(B)})}{\sum_{t=1}^{N^{(ZA)}}w_t(\mathbf{x}^{(A)},\mathbf{x}^{(B)})}\mathrm{N}(\mathbf{Z}^{(A)};\boldsymbol{D}(\mathbf{x}^{(A)},\mathbf{x}^{(B)})\boldsymbol{\beta}_r,\boldsymbol{\Sigma}_r),\end{equation} 
with weights $w_r(\mathbf{x}^{(A)},\mathbf{x}^{(B)})=$
\begin{equation}\label{eqn:weights}
\sum_{h=1}^{N}\pi_h\pi_{rh}^{(ZA)}\left(\sum_{l=1}^{N^{(XA)}}\pi_{lh}^{(XA)}\prod_{j=p_{Ac}+1}^{p_{A}}{\psi}_{lx_{j}^{(A)}}^{(j)}\right)\left(\sum_{s=1}^{N^{(B)}}\pi_{sh}^{(B)}\prod_{j=p_{A}+1}^{p}{\phi}_{lx_{j}^{(B)}}^{(j)}\right).
\end{equation}
Mixtures of linear regressions, even with main effects only in the design matrices, can be
highly flexible globally  \citep{wadedunson}.  Because the weights and means of $\mathbf{Z}^{(A)}$ depend on
$\mathbf{X}$, MM-FC can capture a variety of complex
conditional distributions for $\mathbf{Z}^{(A)}$, even 
relationships beyond those encoded in $\boldsymbol{D}(\mathbf{X})$.  
Nonetheless, in modest sized samples it may be prudent to include interaction terms in
$\boldsymbol{D}(\mathbf{X})$, which can allow the model to use fewer mixture
components and therefore potentially improve inferences.  

The expressions in \eqref{eqn:weightsum} and \eqref{eqn:weights} also 
reveal features about two other possible specifications that use three
sets of clusters. 
Suppose one instead assumes that $\mathbf{Z}^{(A)}$ and $\mathbf{X}$ are locally independent,
   so that $\boldsymbol{D}(\mathbf{X})$ has only an intercept
   term. 
This implies that the mixture of normals for $f(\mathbf{Z}^{(A)}\mid
   \mathbf{X})$ has weights that are dependent on $\mathbf{X}$ but
   means that are not. Several researchers have described the
   downsides of such models \citep[e.g.,][]{dunson:bhat,
     banerjee,deyoreo:binary}, which may not perform well in modest sized samples due
     to the need to introduce many clusters. 
On the other hand, suppose that one allows for local
   dependence between $\mathbf{Z}^{(A)}$ and $\mathbf{X}$, but assumes
   $H^{(ZA)}$, $H^{(XA)}$, and
   $H^{(B)}$ are independent. This yields $f(\mathbf{Z}^{(A)}\mid
   \mathbf{X}=\mathbf{x})=\sum_{l=1}^{N^{(ZA)}}\mathrm{Pr}(H^{(ZA)}=l)\mathrm{N}(\mathbf{Z}^{(A)};\boldsymbol{D}(\mathbf{x})\boldsymbol{\beta}_l,\boldsymbol{\Sigma}_l)$,
   which is a mixture of normal kernels with constant
   weights. As noted by \citet{wadewalker}, in curve fitting
   (regression) contexts, mixtures with constant weights tend to generate
   lower quality predictions than mixtures with covariate dependent
   weights, like those for MM-FC in \eqref{eqn:weights}.
%


The model for $\mathbf{Y}^{(A)}$, obtained by marginalizing over
$\mathbf{Z}^{(A)}$, is a mixture of probit regressions. This has been shown to
be very flexible and able to accommodate complex associations among
ordinal variables, as well as nonstandard regression trends
\citep{deyoreokottas}.

Turning to $\mathbf{X}^{(A)}$ and $\mathbf{X}^{(B)}$, we have
$P(\mathbf{X}^{(A)},\mathbf{X}^{(B)})= $
\begin{equation}\label{eqn:joint_dens_nom}
\sum_{h=1}^N \pi_h  \left(\sum_{l=1}^{N^{(XA)}}\pi_{lh}^{(XA)}\prod_{j=p_{Ac}+1}^{p_{A}}\mathrm{categ}(X_j^{(A)};\boldsymbol{\psi}_l^{(j)})\right)\left(\sum_{s=1}^{N^{(B)}}\pi_{sh}^{(B)}\prod_{j=p_A+1}^{p}\mathrm{categ}(X_j^{(B)};\boldsymbol{\phi}_s^{(j)})\right).
\end{equation}
This can be rewritten as $P(\mathbf{X}^{(A)},\mathbf{X}^{(B)})= $
\begin{equation}\label{eqn:joint_dens_nom_2}
\sum_{h=1}^N \sum_{l=1}^{N^{(XA)}}\sum_{s=1}^{N^{(B)}} \pi_h\pi_{lh}^{(XA)}\pi_{sh}^{(B)}\left(\prod_{j=p_{Ac}+1}^{p_A}\mathrm{categ}(X_j^{(A)};\boldsymbol{\psi}_l^{(j)})\right)\left(\prod_{j=p_A+1}^{p}\mathrm{categ}(X_j^{(B)};\boldsymbol{\phi}_s^{(j)})\right).
\end{equation}
This is a mixture of products of independent multinomial
distributions. Such models have the ability to capture any
multivariate categorical data distribution for large enough numbers of
mixture components \citep{dunsonxing}.  Marginally, we also have  
$P(\mathbf{X}^{(A)})=\sum_{l=1}^{N^{(XA)}}\mathrm{Pr}(H^{(XA)}=l)\prod_{j=p_{Ac}+1}^{p_A}\mathrm{categ}(X_j^{(A)};\boldsymbol{\psi}^{(j)}_{l})$, where 
$\mathrm{Pr}(H^{(XA)}=l)=\sum_{h=1}^N\pi_h\pi_{lh}^{(XA)}$.  
Thus, $\mathbf{X}^{(A)}$ also follows a mixture of
products of multinomials.

From \eqref{eq5} -- \eqref{eq12}, marginalizing over $H_i$ yields
$\mathrm{Pr}(H_i^{(ZA)}=r,H_i^{(XA)}=l,H_i^{(B)}=s)=\sum_{h=1}^N\pi_h\pi_{rh}^{(ZA)}\pi_{lh}^{(XA)}\pi_{sh}^{(B)}$. Thus, although $H_i^{(ZA)}$, $H_i^{(XA)}$, and $H_i^{(B)}$ are independent
conditional on $H_i$, dependence is induced upon marginalization. This
dependence helps the model capture associations among variables in
$\mathbf{Y}^{(A)}$ and $\mathbf{X}^{(A)}$, as well as among the focus
and remainder variables.  The latter associations are strengthened by
the local dependence of $\mathbf{Z}^{(A)}$ on $(\mathbf{X}^{(A)},
\mathbf{X}^{(B)})$ through the regression in \eqref{eqn:data_model},
and the covariate-dependent weights that result from the mixture. 
We note, however, that any marginal dependence of $\mathbf{X}^{(A)}$ 
 with $\mathbf{X}^{(B)}$ has to be captured mostly
 by the TF prior distribution on the components, which suggests that
 these dependencies are the most difficult for the model to capture.


\section{Simulation Studies}\label{sec:simulations}
We conduct a series of simulation studies to investigate the properties of the MM-FC,
especially  in comparison to similar models that do not distinguish focus and remainder variables.  We 
use the MM-FC as an engine for multiple imputation of 
missing data, and assess the
potential benefits of classifying variables with high rates of missingness as focus
variables ${A}$
and other variables as remainder
variables ${B}$. 
We consider eight scenarios defined by a full factorial experiment
with three binary factors: rate of missingness in the focus variables (``high'' is 30\% missing,
``low'' is 5\% missing), number of variables classified as focus 
variables (``few'' is $p_{Ac} = 2$ and $p_A=4$, ``more'' is $p_{Ac}= 4$ and $p_A=8$), and sample size (``small'' is $n=500$, and ``large'' is  
$n=3000$).  Across all scenarios,  $p_{B}=8$, and the remainder variables have $5\%$ missing values.  
When all variables in ${A}$ are missing at a high rate, the probability that a 
given observation is complete is $0.02$ when the number of focus variables $p_{A}=8$ and is $0.12$ when $p_A=4$, essentially prohibiting complete-case analysis.

We generate complete datasets to ensure interaction effects and complex dependencies, 
both among variables within ${A}$ and variables across ${A}$ and  ${B}$. 
The complete data are not generated directly from a MM-FC; rather, we primarily use a series of generalized linear models. 
The data-generating mechanism for $\mathbf{Y}^{(A)}$ includes two and three-way interaction effects, but 
we use a default application of the MM-FC that includes 
only main effect terms in the design vector $\boldsymbol{D}(\cdot)$. See the supplementary material for a detailed description of how the data are generated.

 In each scenario, we repeat the process of generating data and
  randomly deleting values 50 times, using a missing
  completely at random  mechanism. In each dataset, we use the MM-FC  
to generate $m=10$ completed datasets by drawing from the posterior
predictive distribution (which assumes any missing values are MAR).  We run each implementation of the MCMC algorithm 
long enough to obtain $m=10$ sets of imputations for the missing
values, using the completed data set from every $2000$th iteration. We
began saving imputations after discarding $20000$ iterations
as burn-in.

We use the methods of \citet{rubin} for inferences on all marginal
and bivariate probabilities associated with the $p$-way contingency
table. In each completed data set $l$, where $l=1,\dots,m$, let
$q^{(l)}$ be the estimate of a particular cell probability $Q$, and
let $u^{(l)}$ be the estimate of its variance . Here, $q^{(l)}$ is the
empirical proportion of observations in the particular cell, and
$u^{(l)} = q^{(l)}(1-q^{(l)})/n$. To make inferences about $Q$, we use
the point estimate  $\bar{q}_m = \sum_{l=1}^mq^{(l)}/m$ with associated
variance, $T_m=(1+1/m)b_m+\bar{u}_m$, where
$b_m=\sum_{l=1}^m(q^{(l)}-\bar{q}_m)^2/(m-1)$ and
$\bar{u}_m=\sum_{l=1}^mu^{(l)}/m$. Interval estimates are based on
$(\bar{q}_m-Q)\sim t_{\nu_m}(0,T_m)$, where $t_{\nu_m}$ represents a
$t$-distribution with $\nu_m=(m-1)(1+\bar{u}_m/\{(1+1/m)b_m\})^2$
degrees of freedom. See  \citet{reiterrag} for a review of multiple
imputation inference.

\subsection{Performance of MM-FC}\label{sec:evaluate}
Here, we summarize our main findings, focusing on the scenarios with a high rate of missingness among the focus variables as the model is particularly intended for such situations. Details and additional results are in the supplementary material.

\begin{figure}
 \centering
\includegraphics[height=2.6in,width=2.6in]{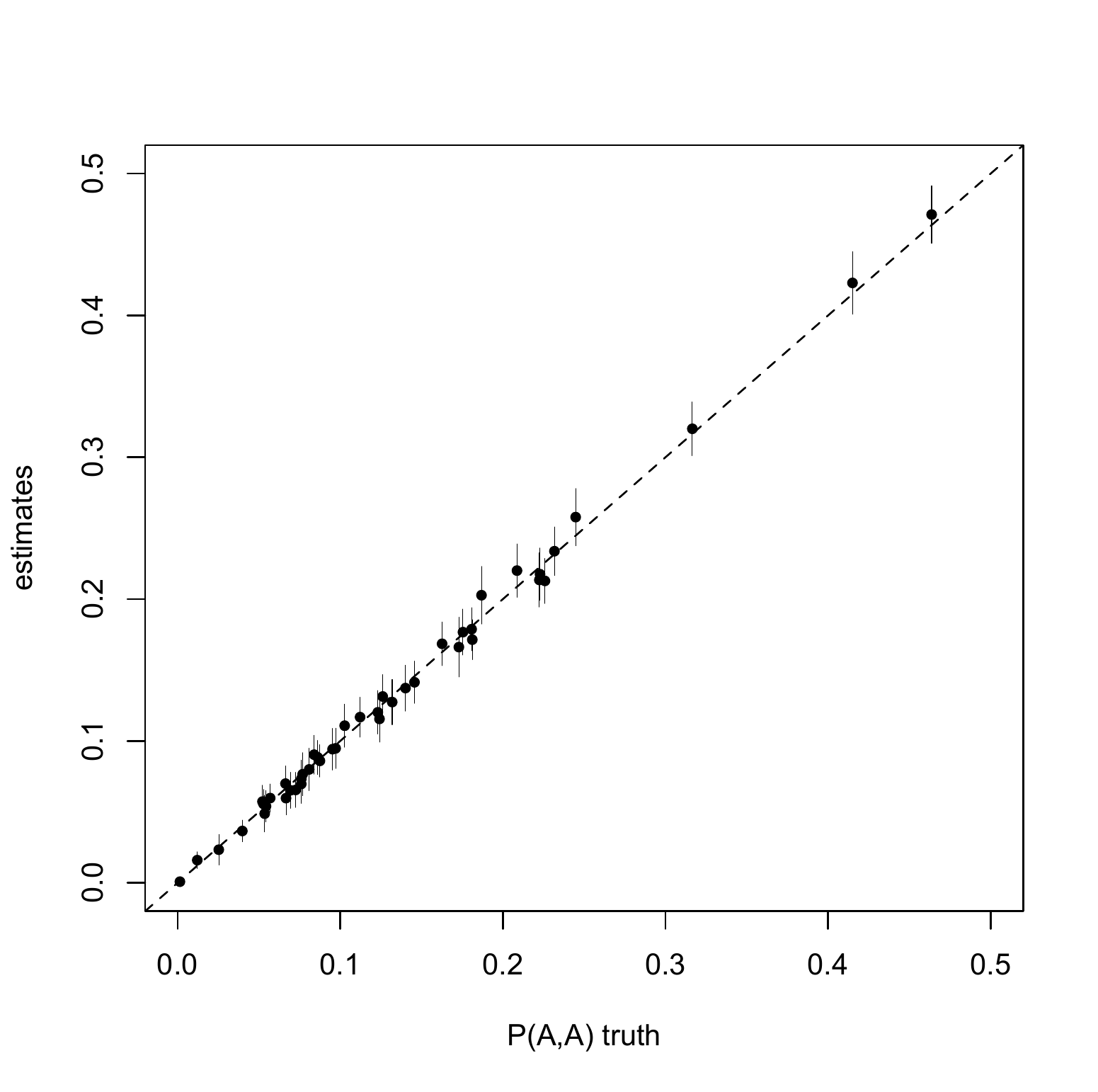}
\includegraphics[height=2.6in,width=2.6in]{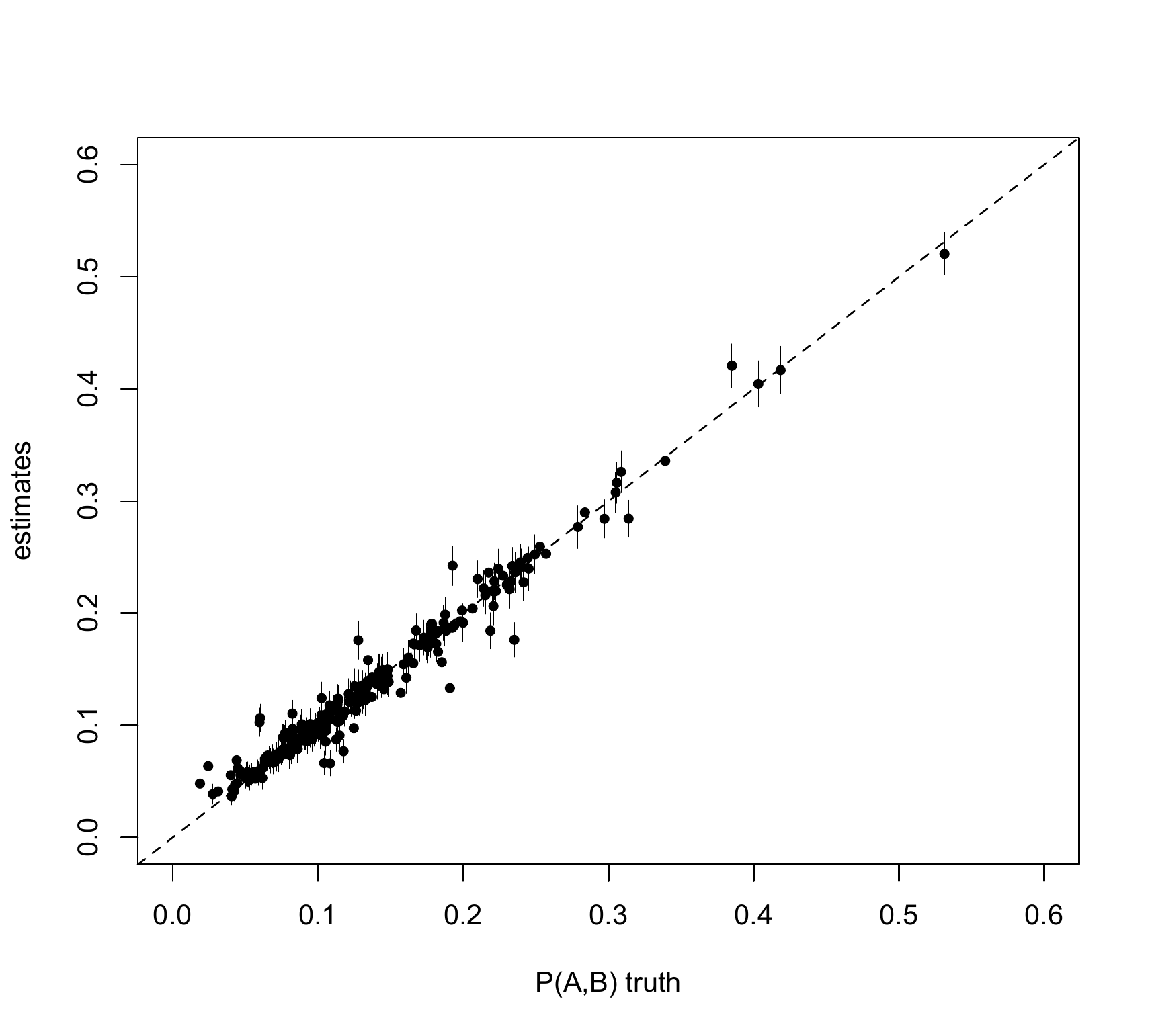}
\caption{Multiple imputation point estimates and 95\% confidence intervals from
  one randomly drawn simulation run with high missingness, few focus
  variables, and large sample size. Left panel includes all bivariate
             probabilities associated with pairs of variables in
             ${A}$, and right panel includes all bivariate
             probabilities for pairs of variables from
             ${A}$ and ${B}$.  Trends are
             similar in other simulation runs for this setting.} 
\label{fig:MI_ests_large}
\end{figure}

As expected, the MM-FC estimates the distribution among the focus
variables well. As an example, Figure \ref{fig:MI_ests_large}
displays the multiple imputation point estimates and $95\%$ confidence
intervals for bivariate probabilities among ${A}$ variables
for one randomly sampled simulation run with $n=3000$ and ``few''
focus variables.  Averaged over simulations, the absolute
errors of the point estimates for the marginal probabilities suggest
the model accurately captures the distribution of $A$ in both the ``few'' and ``more'' settings:
the average across the 11 probabilities is less than $0.009$ when $n=3000$ and less than $0.021$
when $n=500$. 
The same holds for the 45 bivariate probabilities in
${A}$ (these probabilities range from approximately $0.001$ to $0.46$):
the average is less than $0.008$ when $n=3000$ and $0.016$ when $n=500$.
The empirical coverage rates, i.e., the percentage of the fifty
multiple imputation $95\%$ confidence intervals that contain their
corresponding expected values, for the marginal and bivariate
probabilities are generally at or slightly below the nominal $95\%$
level. For example, in the setting with few ${A}$ variables and large sample size, the
average of the $56$ empirical coverage rates is $0.93$. A handful of rates for
bivariate probabilities fall between   
$80\%$ and $90\%$. One interval corresponding to a bivariate
probability between two ordinal variables has low coverage at around
$65\%$.  This estimand corresponds to a small probability of $0.012$. The mean absolute error of the point estimate is only $0.0056$, 
which is a typical value among the $45$ bivariate probabilities.

\begin{figure}[t]
 \centering
\includegraphics[height=3in,width=3in]{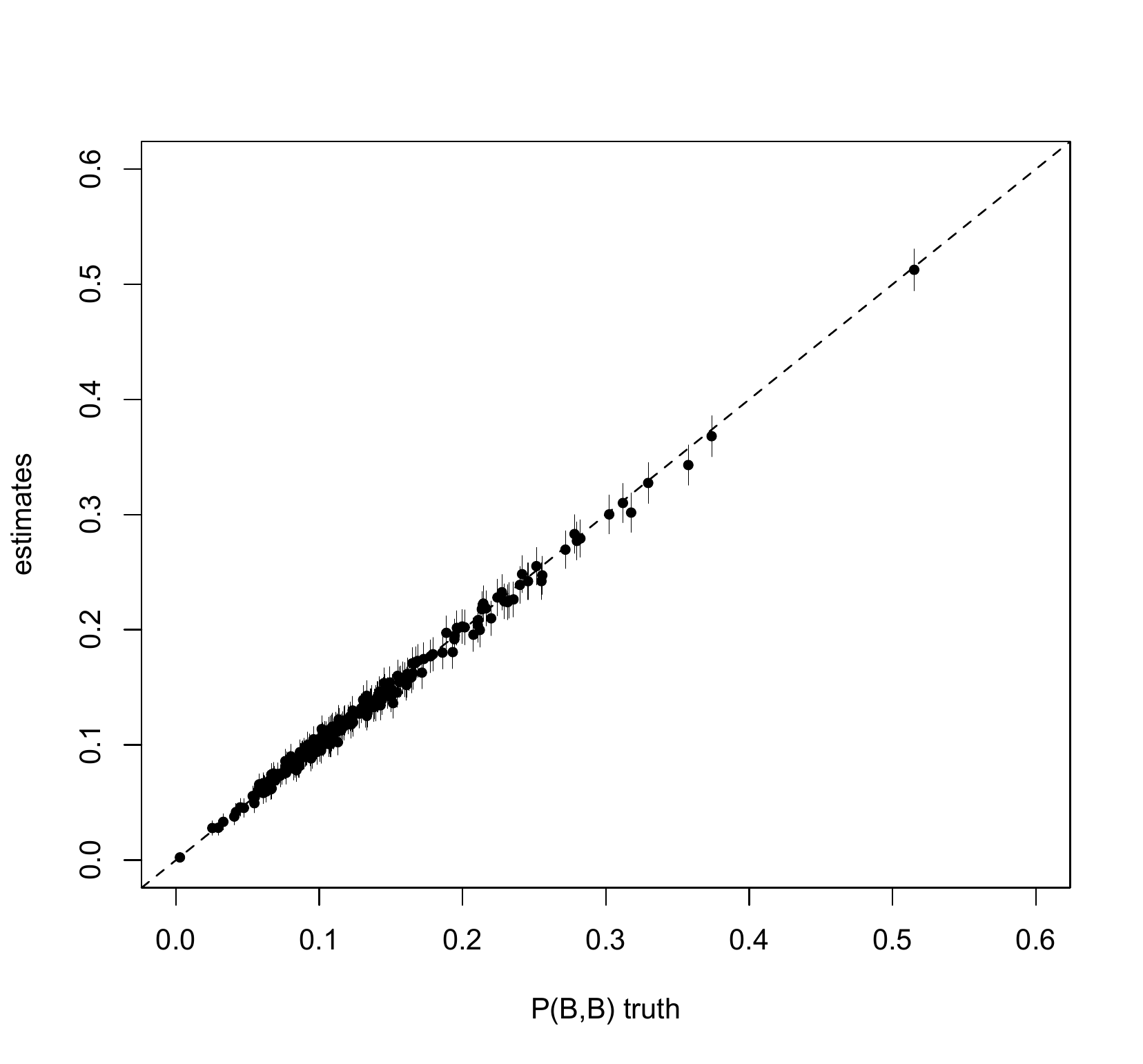}
\caption{Multiple imputation point estimates and 95\% confidence
  intervals for all bivariate probabilities for pairs of variables in
  ${B}$ from  one randomly drawn simulation run with high missingness, few focus
  variables, and large sample size.} 
\label{fig:MI_ests_large_B}
\end{figure}

The MM-FC also generates reliable inferences for the distributions
among the remainder variables, as is clear from Figure \ref{fig:MI_ests_large_B}. This is not surprising, since we only impute a small fraction of
missing values. In the setting with few ${A}$ variables and large sample size, the average of the $233$ empirical coverage rates for the marginal and bivariate probabilities is approximately $0.98$.

Most interesting is the performance of MM-FC for estimating
relationships between focus and remainder variables.  In general, the
model continues to offer estimates with modest absolute errors: the mean absolute errors of the  probabilities for pairs of variables from ${A}$ and ${B}$ are less than $0.011$ when $n=3000$ and less than $0.016$ when $n=500$. 
The empirical coverage rates are around $76\%$ to $79\%$ in both settings with large
sample size, and $90\%$ in both settings with small sample size.  Evidently, in the simulations with large sample size, the
modest biases resulting from MM-FC are large enough relative to the standard errors to
reduce coverage rates, whereas this is not the case with the small sample size. 
As predicted, the model is least accurate when estimating relationships between nominal variables in ${A}$ and variables in 
${B}$. For instance, in the setting with more focus variables and large sample size, the average coverage rate when $A$ and $B$ are both ordinal is $86\%$, 
the average coverage rate when $A$ is ordinal and $B$ is nominal is
$82\%$, the average coverage rate when $A$ is nominal and $B$ is
ordinal is $77\%$, and the average coverage rate when $A$ and $B$ are
both nominal is $57\%$.

Looking across all eight scenarios, in general the model performs more effectively with low
fractions of missing data in both the focus and remainder variables.
As would be expected, increased sample size results in better ability
to capture relationships among the variables and hence lower absolute
errors. For a given sample size and rate of missingness, the
differences in performance arising from few versus more focus
variables are not significant.

\subsection{Evaluation of Use of Focus Variables}\label{sec:compare}

The simulations in Section \ref{sec:evaluate} suggest that the MM-FC
does what is intended: use separate clusters for focus variables to fit
their distribution accurately, possibly at the expense of accurately
modeling remainder variables.  The question now is whether or not
the MM-FC offers gains over models that do not distinguish
between focus and remainder variables.  To examine this, we compare
the MM-FC against MM-Mix.
We use MM-Mix to
generate $m=10$ completed datasets for the same fifty simulations used
in Section \ref{sec:evaluate}.

In each simulation run, after generating ten completed datasets from each model, we compute the Hellinger distance between the estimated and true joint distribution of $P(A)$ in each completed dataset.  We use 
only cells for which the true probability is
at least $8\times 10^{-6}$. We then average the Hellinger distances across the $10$ completed datasets.
We do this also for $P({B})$
and $P({A},{B})$. In every scenario, the Hellinger distances for $P({A})$ are smaller
under MM-FC than MM-Mix. In the scenario with high rate of missingness, large $n$, and few
${A}$ variables, on average the Hellinger distance for MM-FC 
is about $50\%$ smaller than that for MM-Mix. The
Hellinger distances for ${B}$ are similar for both models on average;
however, the distances for MM-FC have much smaller variance across
the simulations than those for MM-Mix, indicating that the MM-FC is more
stable in offering a high quality estimate of the distribution of
${B}$.  MM-Mix produces Hellinger distances for 
$P({A},{B})$ that are slightly smaller than
those produced by MM-FC, indicating that some strength of
dependence between ${A}$ and ${B}$ is lost by
the introduction of separate but dependent cluster assignments. The
differences between the models are in general more pronounced when the
dimension of ${A}$ is smaller than that of
${B}$, i.e., under the few focus variables setting.

\begin{figure}
 \centering
\includegraphics[height=2.6in,width=2.6in]{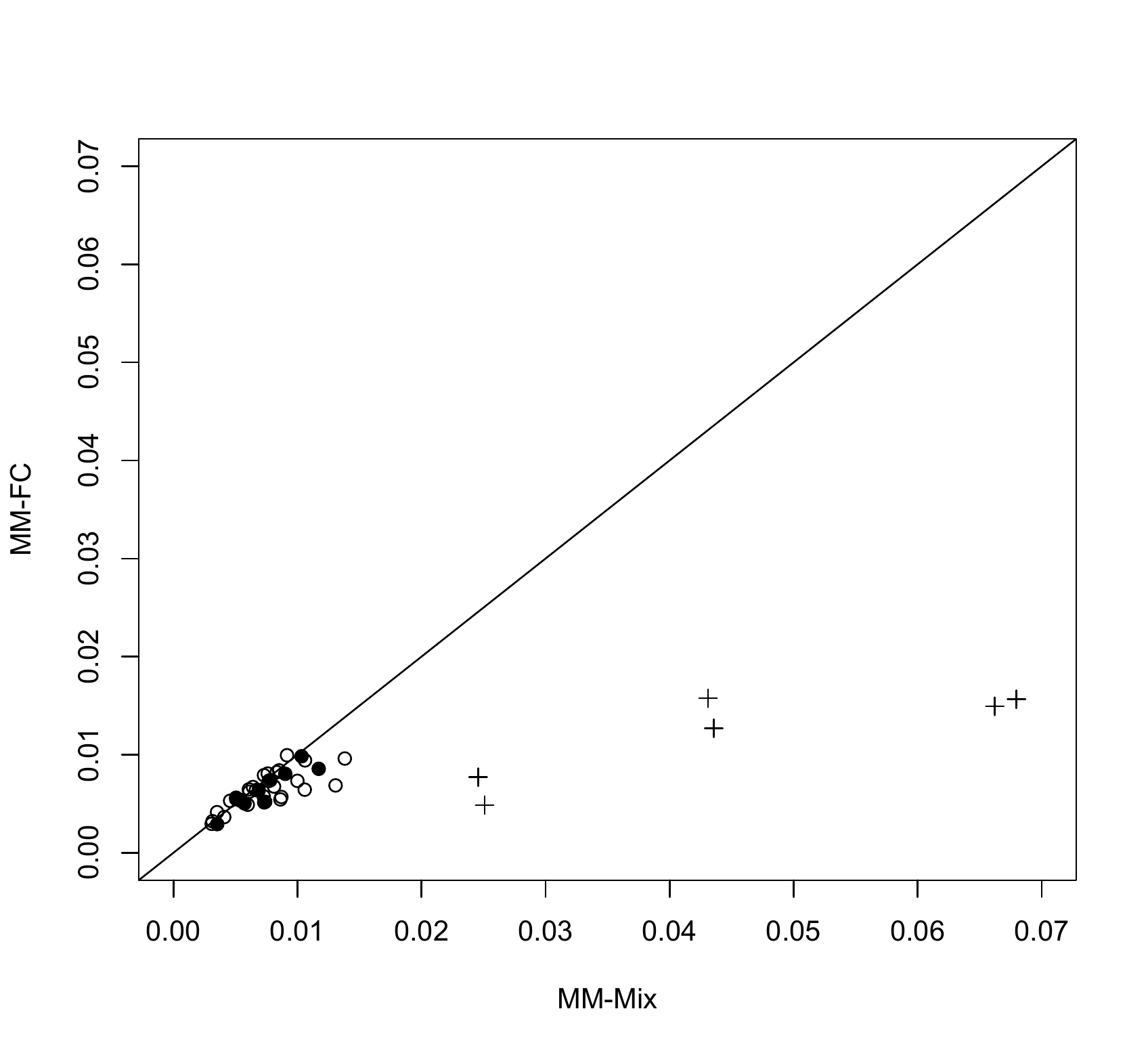}
\includegraphics[height=2.6in,width=2.6in]{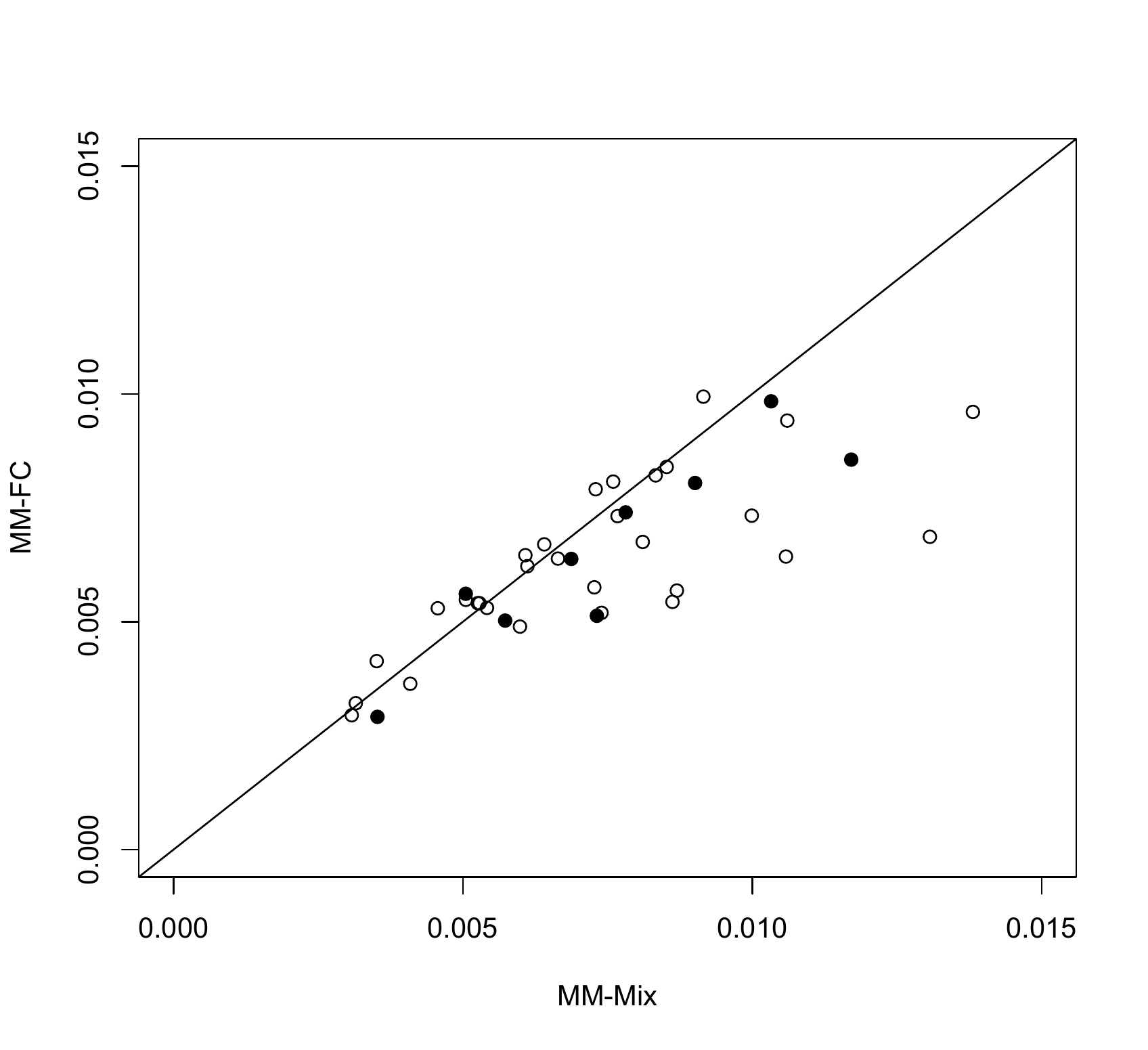}
\caption{Simulation setting with high missing rate, few focus variables, and large sample size. Left: Absolute errors of the 45 bivariate probabilities associated with all pairs of ${A}$ variables averaged over 50 simulations from MM-FC versus MM-Mix. Solid circle indicates ordinal-ordinal probabilities, $+$ indicates nominal-nominal probabilities, and open circles represent ordinal-nominal probabilities. Right: Close up of the lower left part of this figure.}
\label{fig:hfl_AE}
\end{figure}

\begin{figure}
 \centering
\includegraphics[height=2.6 in,width=2.6in]{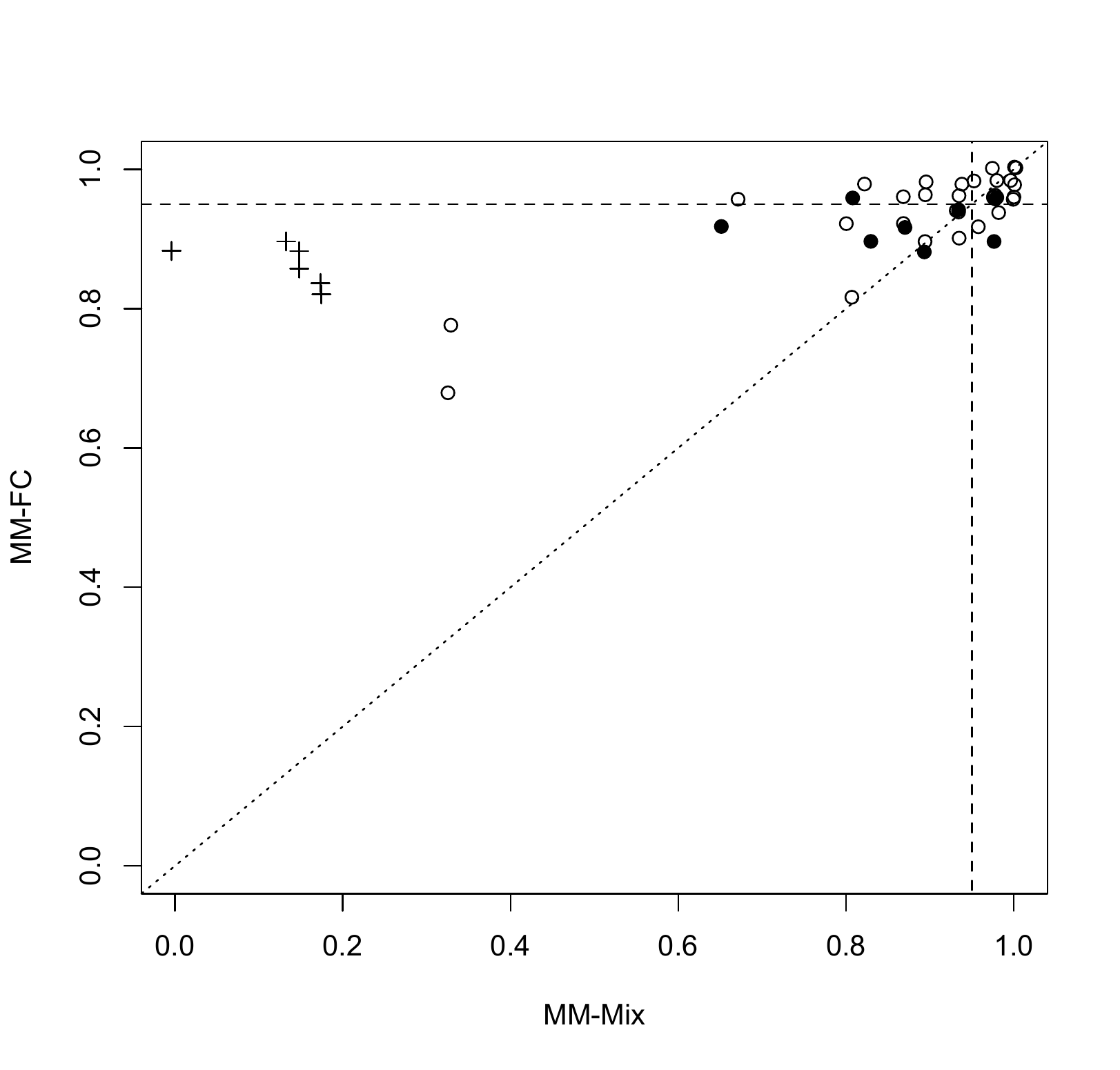}
\caption{Simulation setting with high missing rate, few focus variables, and large sample size. Coverage of $95\%$ confidence intervals for the $45$ estimands involving bivariate probabilities associated with pairs of ${A}$ variables from MM-FC versus MM-Mix. Points have been jittered for readability.}
\label{fig:hfl_cov}
\end{figure}

In all eight settings considered, the mean absolute errors resulting
from estimates of bivariate probabilities among ${A}$ are lower under MM-FC
than MM-Mix, with differences too large to be plausibly explained by Monte
Carlo error. Additionally, the empirical coverage rates from MM-FC
are closer to the nominal rate of $95\%$ than those from MM-Mix,
which are often lower. In particular, MM-Mix often is less accurate
for nominal-nominal relationships within ${A}$, as
illustrated in Figure \ref{fig:hfl_AE}. This is likely
because of the assumption of a common latent class for
$\mathbf{X}^{(A)}$ and $\mathbf{X}^{(B)}$. 
The estimated distribution of $\mathbf{X}^{(A)}$ is 
degraded by having to estimate with common clustering the distribution of $\mathbf{X}^{(B)}$, which is of larger
dimension and contains more information due to a smaller rate of
missingness.  The overall difference is not due to one or two quantities
being inaccurately estimated under MM-Mix; as evident in Figure
\ref{fig:hfl_AE} most errors tend to be larger under MM-Mix than
MM-FC.  As evident in Figure \ref{fig:hfl_cov},
coverage rates under MM-Mix average $0.79$ and under MM-FC average $0.93$. 
We also note that in  settings occurring with more focus variables, MM-Mix
is noticeably less accurate on ordinal-ordinal
relationships as well. For figures illustrating these findings, as well as results from all simulations, see the supplementary material.

\begin{figure}[t]
 \centering
\includegraphics[height=4in,width=4in]{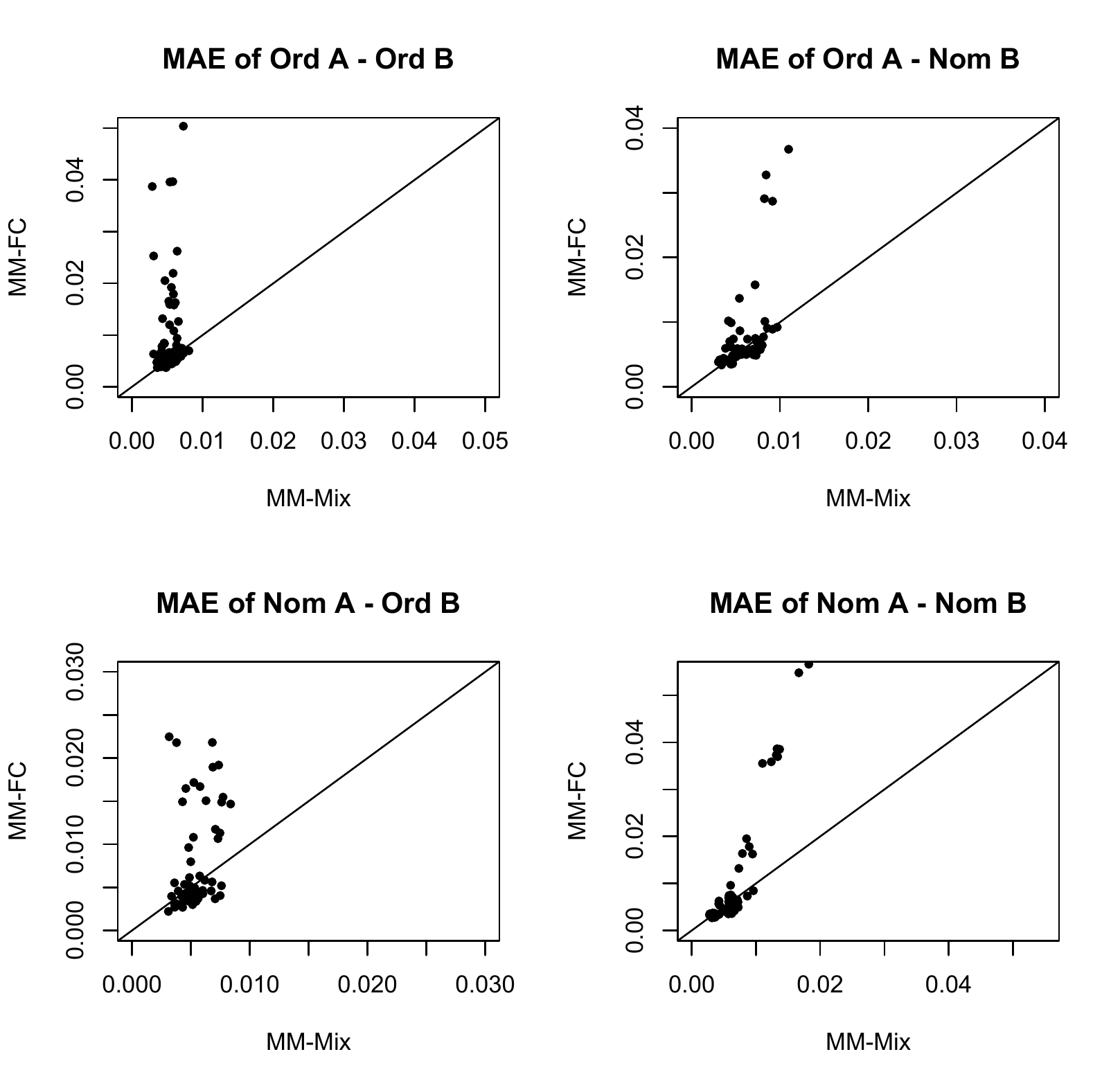}
\caption{Simulation setting with high missing rate, few focus variables, large
sample size. Mean absolute errors of bivariate probabilities for pairs of ${A}$ and ${B}$ variables, organized by variable types.}
\label{fig:hfl_AB}
\end{figure}

Turning to relationships among $A$ and $B$, the simulations suggest mixed results.  For many bivariate probabilities, 
MM-FC and MM-Mix result in similar levels of accuracy. However, MM-FC results in noticeably 
larger errors for some bivariate probabilities. This pattern is exemplified in Figure \ref{fig:hfl_AB}, which 
displays average margins of error for the setting with few focus variables and large sample size. 
We note that the simulation setting in Figure \ref{fig:hfl_AB} is least favorable to MM-FC among all we investigated.   
 Apparently, this scenario has few enough variables overall and
  large enough sample size that MM-Mix is able to characterize the
  joint distribution reasonably well, making the MM-FC look
  comparatively worse on relationships between $A$ and $B$. 
Because the MM-FC introduces more cluster variables, there is a further degree of separation between variables in $A$ and $B$. This improves inference for $A$, however it sacrifices some of the dependence between $A$ and $B$. This therefore leads to some relationships between $A$ and $B$ being captured relatively poorly by MM-FC, hence the relatively large errors for some of the cells, as depicted in Figure \ref{fig:hfl_AB}.
 In other scenarios, the  MM-FC produces fewer relatively
  large errors than are evident in Figure \ref{fig:hfl_AB}. 

Considering the results of all eight simulation scenarios, we find
that MM-FC always captures the distribution of ${A}$ more
effectively than does MM-Mix.  The advantages are especially
evident for nominal-nominal relationships within ${A}$.  
These advantages are greater when the number of remainder variables exceeds the number of focus
variables. The MM-FC also tends to capture the distribution within
  $B$ more effectively than does MM-Mix.  However, MM-FC is
  generally less effective than MM-Mix at capturing relationships
  between nominal $A$ variables and ordinal $B$ variables. For other
  relationships involving $A$ and $B$, the average absolute errors for
  joint probabilities typically are smaller for MM-FC than for
  MM-Mix; however,  MM-FC typically results in
  more probabilities with relatively large errors than MM-Mix does.
In general, the differences between MM-FC and MM-Mix that we have described are more pronounced with a large sample size.


\section{American National Election Survey Analysis}\label{sec:ANES}

\subsection{Data and Modeling Approach}
The ANES has been conducted during presidential election years since $1948$.
 The most recent in this series took place in 2012. 
We work with the data obtained from face-to-face interviews conducted in the two months preceding the presidential election. 
The questionnaire consisted almost entirely of ordered and unordered categorical data, and the median survey length was 90 minutes.

As with many analyses in political science, we are especially interested in measures related to voting behavior, ideology and candidate preference. 
Unfortunately, many of these measures suffer from a high rate of item nonresponse or were not collected for many individuals. For
 instance, liberal-conservative ideology (on an ordered 7 point scale) is missing at a rate of $28\%$, candidate preference in 
$2008$ is missing at a rate of $35\%$, and Tea Party support is
missing at a rate of $17\%$. Only $333$ out of $n=2054$ individuals
have complete data. Most other variables of interest are missing at low rates.

We assume the data are MAR, which is the standard assumption within political science
  \citep{Honaker}.  The limited research explicitly evaluating the MAR
  assumption for unit and item nonresponse in previous ANES studies
  finds some concerns about bias in measures related to voter turnout
  and other outcomes with socially-desirable responses (e.g., racial
  attitudes) but finds little evidence of bias in measures related to
  candidate preference, as is the topic of study here
  \citep{peress2010correcting,bartels1999panel,berinsky2004silent}.

We estimate the MM-FC on the $20$ variables described in Table
\ref{table:variables}. Since Tea Party
support, ideology, candidate preference in 2008, defense spending and
congressional approval are missing at high rates, and most are
important for inference, we include these variables in
${A}$.    
We also include party affiliation and candidate preference in 2012 in
${A}$ because they are substantively important measures for
our analysis. 
We consider all demographic variables and other attitudinal variables
as ${B}$ variables. Thus, we have four ordinal
${A}$ variables, three nominal ${A}$ variables,
eight ordinal ${B}$ variables, and five nominal
${B}$ variables.  We generate $m = 10$ completed datasets, 
using every $5000$th draw from the completed datasets generated by the
MM-FC, after discarding the first $20000$ iterations as burn-in.

The survey includes weights that account for the two-stage stratified
cluster sampling design and post-stratification adjustments.
We do not consider the weights when estimating the MM-FC. A variety
of exploratory data analyses (based on regressing each outcome on the
weights and other variables) suggest that the weights are not
important for predicting any of the variables when the other variables
in Table \ref{table:variables} are in the model. However, we use survey-weighted inference for finite population
quantities after creating the multiple imputations.

\begin{table}
\centering
\begin{tabular}{ l || c | c | c | c }
 Variable & Group & Type & Levels & Percent missing\\
  \hline                   
  \hline
  Party affiliation & $A$ & nominal & 3 & $1$ \\
  Candidate pref. 2012 & $A$ & nominal & 4 & $2$ \\ 
  Candidate pref. 2008 & $A$ & nominal & 3 & $36$ \\ 
  Tea Party support & $A$ & ordinal & 7 & $17$\\
  Ideology  & $A$ & ordinal & 7 & $29$\\
  Defense spending & $A$ & ordinal & 4 & $20$\\
  Congress approval & $A$ & ordinal & 4 & $17$\\
  Democrat approval & $B$ & nominal & 2 & $2$ \\
  Republican approval & $B$ & nominal & 2 & $3$ \\
  Country on track & $B$ & nominal & 2 & $5$\\
  Race & $B$ & nominal & 4 & $0.4$\\
  Gender & $B$ & nominal & 2 & $0$ \\
  Pres. approval & $B$ & ordinal & 4 & $5$\\
  Foreign approval & $B$ & ordinal & 4 & $11$\\
  Health care & $B$ & ordinal & 4 & $7$\\
  Gun importance & $B$ & ordinal & 5 & $0.4$\\
  Social security spending & $B$ & ordinal & 3 & $3$\\
  Education & $B$ & ordinal & 5 & $0.8$\\
  Age & $B$ & ordinal & 6 & $3$\\
\end{tabular}
\caption{Summary of the variables included in the joint model of the
  ANES data.}
\label{table:variables}
\end{table}

\subsection{Analysis Results}
Conducted during political campaigns, pre-election surveys are especially  
concerned with identifying the subset of the electorate that will
actually vote and with predicting the preferences of voters who
are undecided between the candidates. Thus, our analysis focuses on
candidate preference (vote intent) in 2012.  We start by looking at
how candidate preference relates to two of the variables with high
rates of missingness: candidate preference in 2008 and ideology.
Candidate preference in 2008 is likely missing at a high rate due to
recall issues, lack of eligibility, and the fact that not all
respondents were asked who they preferred in the previous
election. Ideology ranges from very liberal to very conservative on a
$7$ point ordinal scale. Item nonresponse on the ideology question tends to reflect respondent difficulty in using the scale to capture ideological preferences or perceived sensitivity in answering the question \citep{treier2009}.

\begin{figure}
\centering
\includegraphics[height=2.1in, width=5.3in]{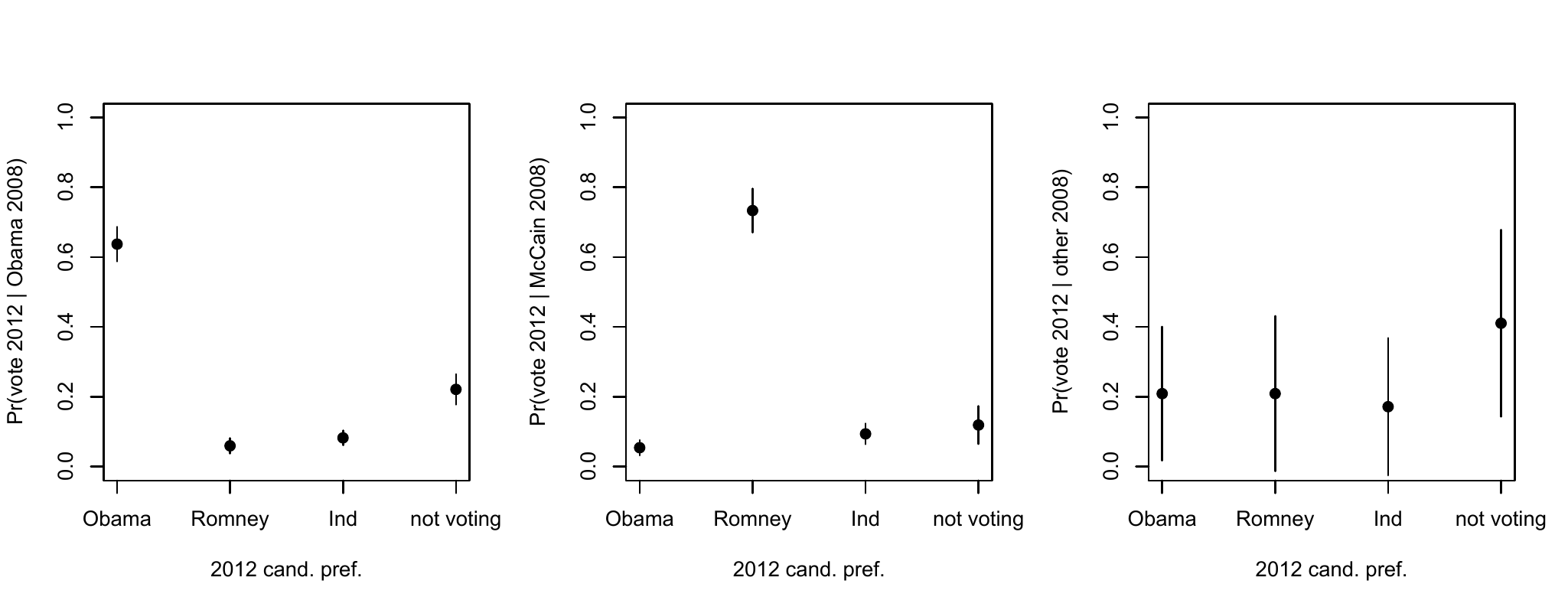}
\caption{Point estimates and $95\%$ uncertainty bands for candidate preference in 2012 conditional on 2008 candidate preference of Obama (left), McCain (middle) and other (right).}
\label{fig:2012-2008-vote}
\end{figure}

Figure \ref{fig:2012-2008-vote} describes the 
relationship between candidate preference in 2012 and 2008 based on the
multiply-imputed datasets. Once we account for missingness, we find that
only $64\%$ of those who preferred Obama in 2008 intend to vote for
him again in 2012, with a significant proportion saying they will not
vote. Similarly, those who preferred McCain in 2008 report they plan
to vote for Romney in 2012; however, this probability is larger by
about $10\%$. In other words, there was greater stability in
preferences across elections on the Republican side than on the Democratic side. We
also obtain estimates for candidate preference in 2012 as a function
of ideology. Although not shown, we find that those most likely to say
they are not voting are those who are liberal-moderate and moderate. Moderate individuals are also most likely to be
undecided in 2012.  

A key substantive question of the 2012 campaign was whether or not Obama could hold
on to Independents between 2008 and 2012. That is, what does Pr(vote
2012 $\mid$ party=Ind, 2008 pref. = Obama) look like? We find that the majority of independents who preferred Obama in 2008 intended to vote for him again in 2012. However, the proportion that said they were not going to vote was larger than the proportion that planned to vote for Romney. That is, Obama did not appear to lose many Independents to Romney, but instead
many of them planned to stay home in the 2012 election.

 For each of the $m=10$ completed data sets, we fit a logistic regression with vote intent in 2012 as the
binary response, indicating whether or not one intends to vote for
Obama. The explanatory variables include main effects and all two-way
interactions for candidate preference in 2008, party, ideology
(liberal, moderate, conservative), and opinion on the Tea Party
(oppose, no opinion, support), all of which are considered
important predictors of vote choice \citep{pasek2009determinants}. 
Ideology is not significant in explaining the way one intends to vote
and is also moderately correlated with 2008 preference and Tea Party
support. We therefore remove this variable from the regression.  
  We combine the ten resulting point and variance estimates of the
  regression coefficients using multiple imputation inference; these
  estimates are given in the online supplement.

\begin{figure}
\centering
\includegraphics[height=2.2in, width=5.3in]{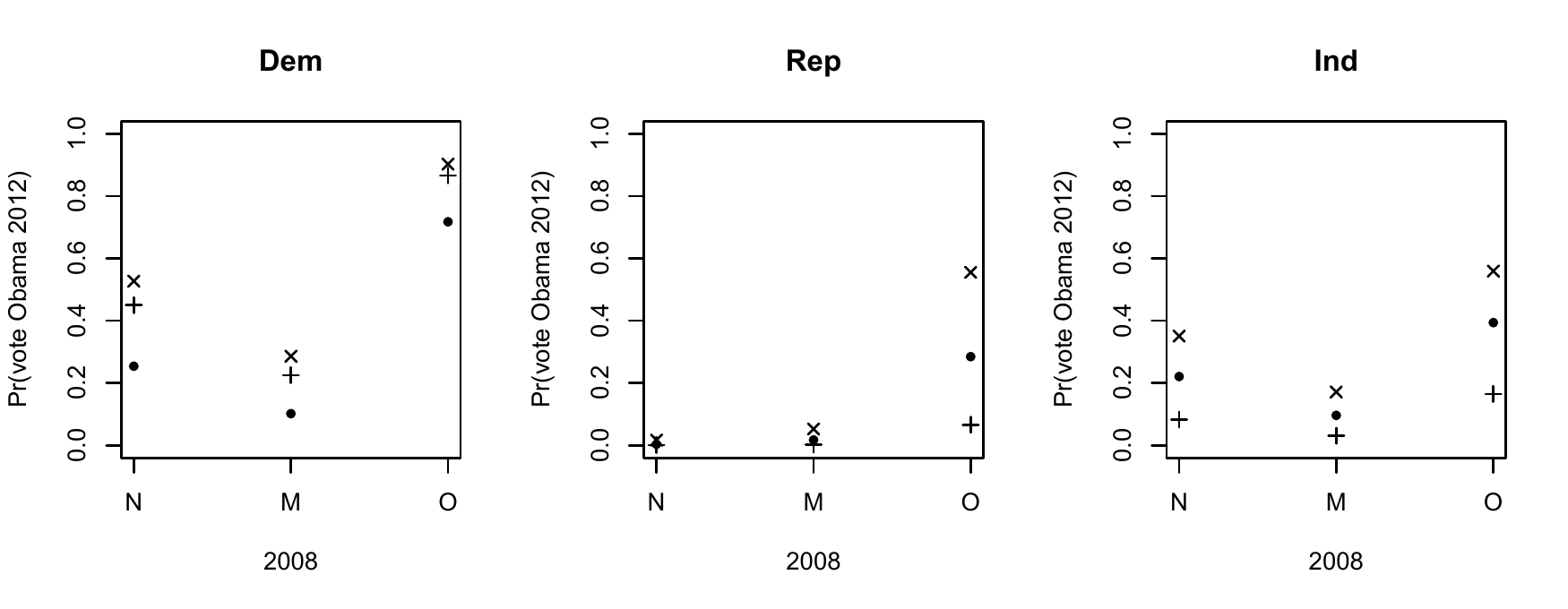}

\caption{Predicted probability of preferring Obama in 2012 for Democrats (left), Republicans (middle) and Independents (right), by 2008 candidate preference (N=neither, M=McCain, O=Obama) and Tea Party support (x=oppose, $+$=support, and $\bullet$=neither).
}
\label{fig:predict}
\end{figure}

While overall the variables are related to 2012 candidate preference in
expected ways, the interaction effects reveal interesting insights about
voter decision making. 
There are significant interactions between party identification (Democrat,
Republican, Independent) and Tea Party support, as well as party and
2008 preference. To visualize and interpret these effects, in Figure
\ref{fig:predict} we plot predicted probabilities of voting for Obama for
each of the $27$ possible combinations of 2008 candidate preference,
party, and Tea Party support. 
Tea Party support is not strongly related to 2012 candidate preference for partisans who
previously voted along party lines: Obama Democrats and McCain
Republicans.  However, Tea Party support is predictive of 2012 vote among Obama
Republicans and
Independents. Opinions about the Tea Party are irrelevant for Democrats -- party loyalty and past support trumps Tea Party opinions.

The analysis also reveals two different types of Independents, with
different strategic implications for the candidates. The first are
those who behave very much like partisan identifiers. Those who claim
to be Independent but support the Tea Party and preferred McCain in
2012 are extremely unlikely to vote for Obama in $2012$, behaving much like self-identified Republicans. Additionally,
Independents who oppose the Tea Party and preferred Obama in $2008$
look very much like self-reported Democrats. This group of
Independents are often called ``closet partisans'' and are not really
``up for grabs'' in the campaign. In contrast, the Independents who are
actually ``in play'' in the election are those who are ambivalent or
cross-pressured.  For example, these include self-reported
Independents who voted for Obama in 2008 but also support the Tea
Party, or who voted for McCain but oppose the Tea Party. This group of
Independents falls in the middle in terms of the probability of preferring
Obama in 2012. 

Another interesting pattern is that Tea Party support
does seem to be important when considering those who are
cross-pressured, i.e., Republicans who preferred Obama in 2008 and
oppose the Tea Party are much more likely to vote for Obama than
Republicans who preferred Obama in 2008 and support the Tea Party. Of
those who supported Obama in 2008, those who support the Tea Party are most likely to vote against him
in 2012.  

For readability, Figure \ref{fig:predict} displays only point
estimates. The uncertainty bands corresponding to
the four largest probabilities as well as most of the small probabilities are
narrow, in that most $95\%$ interval bands have
width less than $0.1$. The uncertainty bands associated with those
who preferred ``neither'' 2008 candidate are often extremely wide. In
particular, uncertainty is largest for Democrats who preferred a
candidate other than Obama or McCain in 2008. 

\begin{figure}[t]
\centering
\includegraphics[height=3.5in, width=4in]{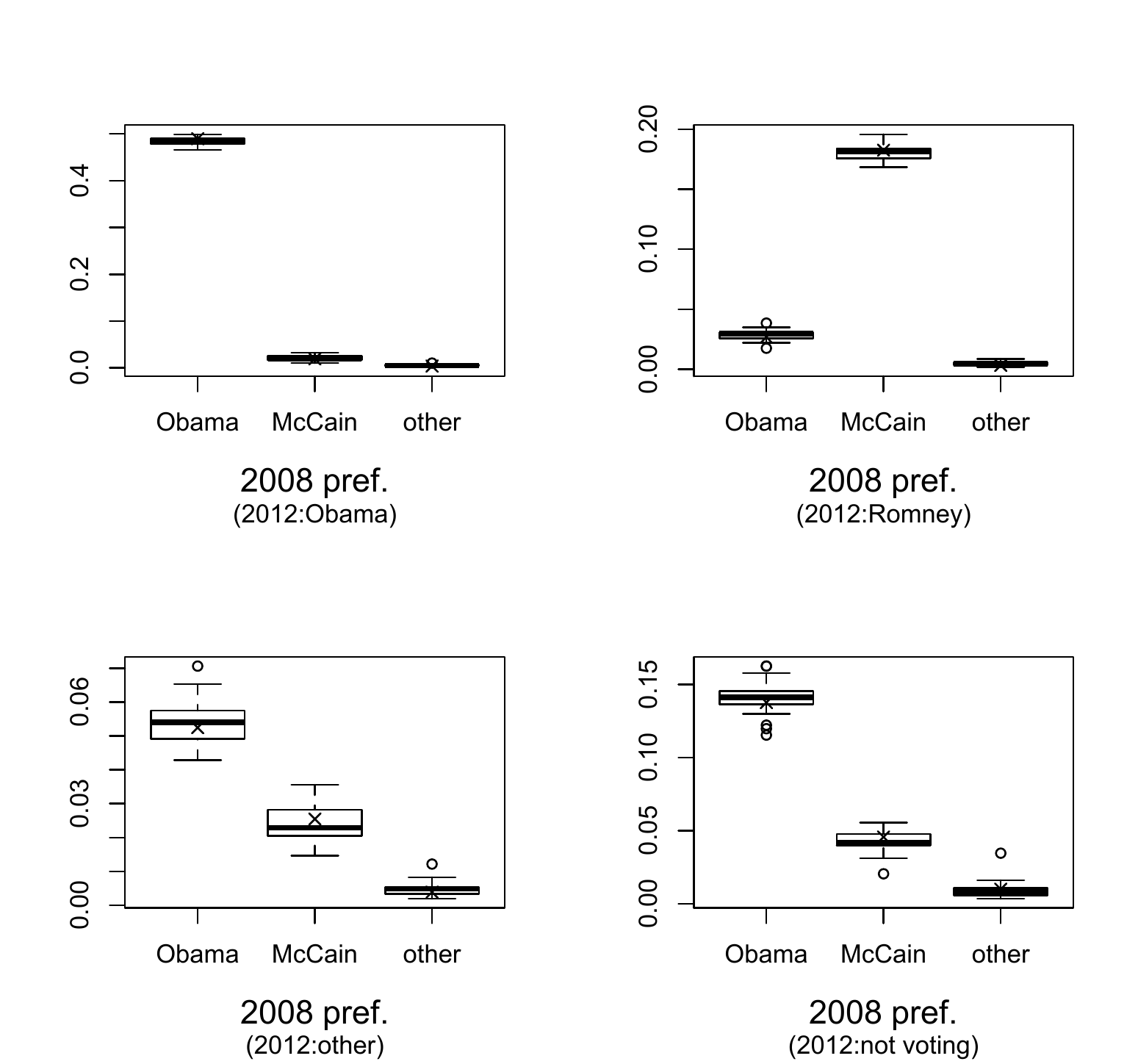}
\caption{Distribution based on replicated data sets for the bivariate distribution of candidate preference in 2012 and candidate preference in 2008 versus point estimate from the multiple completed data sets (x symbol). For instance, the upper left plot gives Pr(Obama 2012, Obama 2008)$\approx 0.5$, while Pr(Obama 2012, McCain 2008) as well as Pr(Obama 2012, other 2008) are both close to zero.}
\label{fig:modelcheck}

\end{figure}

\begin{figure}
\centering
\includegraphics[height=4.5in, width=5in]{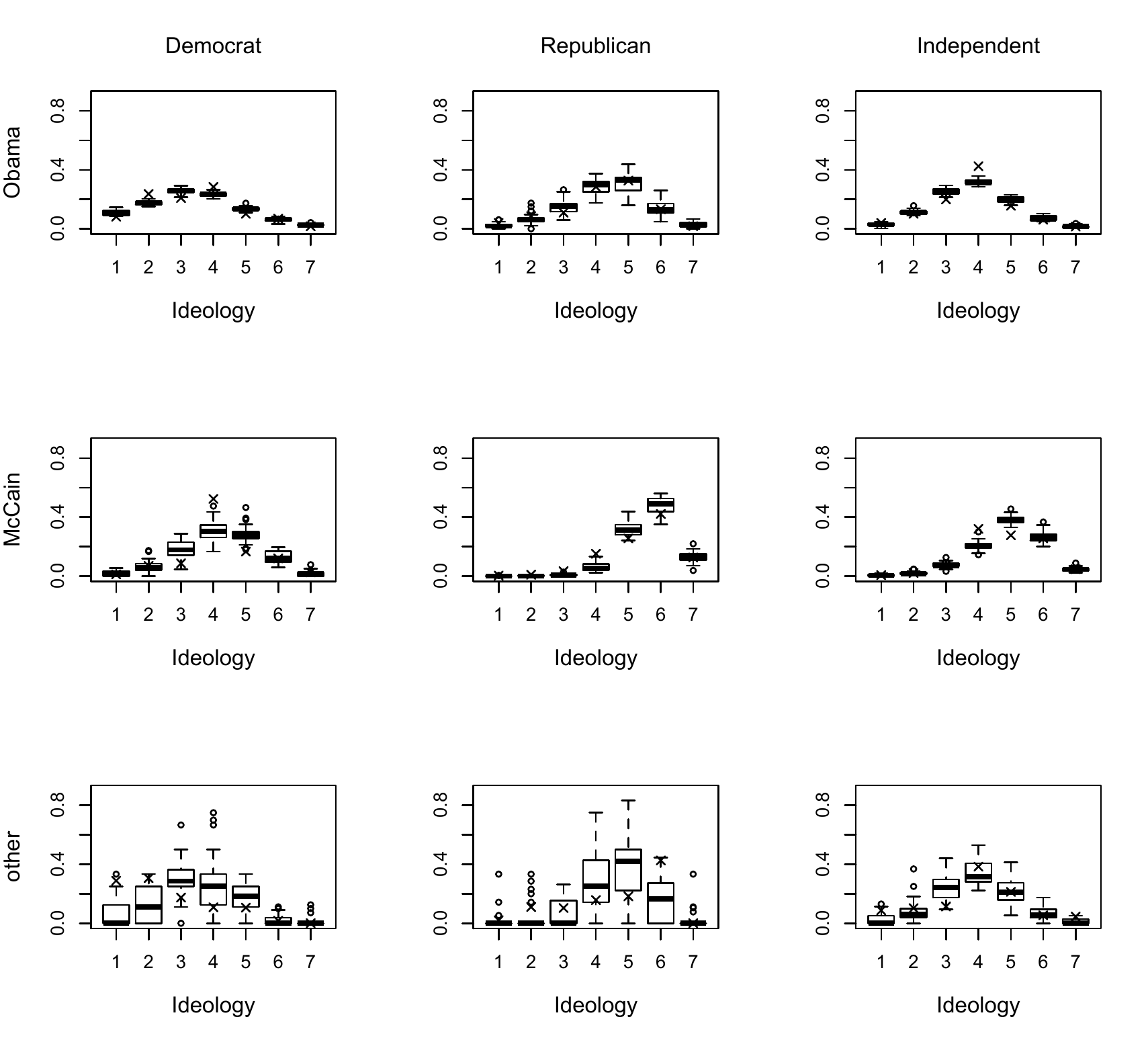}
\caption{Distributions based on replicated data sets for the distribution of ideology conditional on candidate preference in 2008 (indicated by row labels) and party (indicated by column labels) versus point estimate from the multiple completed data sets (x symbol).}
\label{fig:main-effects-check}

\end{figure}

To check the plausibility of the imputations generated by the MM-FC,
we follow the advice in \citet{abayomi} and
\citet{gelman-model-checking} by comparing distributions of imputed
and observed values.  These distributions  
exhibit similar patterns with only slight differences, suggesting the
imputations are plausible.  
We also evaluate the plausibility of the MM-FC imputations with posterior predictive checks \citep{zaslavsky}. Using 25 draws of the parameters from the
posterior distribution, we generate 25 replicated datasets,  compute
statistics of interest with the replicated data, and compare the 
distribution of these statistics with the corresponding values
computed with the $m=10$ multiple imputations.  
We choose statistics that
correspond to inferences of substantive interest. As examples, Figure
\ref{fig:modelcheck} displays bivariate distributions of vote intent
and candidate preference, and 
Figure \ref{fig:main-effects-check} displays distributions of ideology
conditional on candidate preference and party. There are no obvious
indications that the MM-FC generates implausible
imputations. Figure \ref{fig:main-effects-check}, which involves
$\mathbf{Y}^{(A)}$ and pairs of variables in $\mathbf{X}$ variables, provides
some assurance that the specification for $\boldsymbol{D}(\mathbf{X})$ is
reasonable for these data.
We include additional posterior predictive checks in the supplementary material.

As comparisons, we implemented the chained equations approach to multiple imputation
\citep{raghunathan2001multivariate, vanbuuren2011} using default specifications in
the MICE software in R
\citep{vanbuuren2011}.  We also implemented the standard method used in
political science using default specifications in the software package ``Amelia''
\citep{Honaker}, which 
generates discrete-valued imputations via transformations and
rounding of draws from a multivariate normal distribution. For these approaches, the posterior predictive checks
indicated serious inadequacies in model fit. See the
supplemental material for these results. 

\section{Discussion}\label{sec:discussion}

The simulations indicate that separating variables into focus
and remainder variables can result in improvements in estimation
accuracy. These gains are manifested most clearly for the focus
variables.  There also can be gains when estimating relationships
between ordinal focus variables and the remainder variables.  However,
the separation comes at a cost for estimating relationships between
nominal focus variables and the remainder variables. We note that we also observed improvements in accuracy when evaluating focused clustering when using
independent multinomial product kernels for both sub-models; see the
supplementary material for details.

These findings suggest future research directions around tailoring the
selection of the focus variable set.  
For example, there may be advantages to including in the focus variable set all
variables that are of primary interest, even when they have small
rates of missing values.  This can allow the model to concentrate its
fitting power on the joint distributions of the variables of interest,
but still use the remainder variables to improve imputations.  As
another option, the analyst might include variables that are not of
direct interest, or are observed with low rates of missingness, but
are highly correlated with key focus variables.  Finally, the results
of these investigations suggest extending the MM-FC to
 allow the data to determine automatically the most beneficial
allocations to focus and remainder variables.  


As indicated in Section  \ref{properties}, various marginal
distributions of the MM-FC are known to possess desirable 
properties (when the number of clusters is allowed to be infinite), such as large support and consistency. Large support refers to the ability of the prior model to generate distributions that are arbitrarily close to any true data-generating distribution. \citet{dunsonxing} established that the marginal model for $P(\mathbf{X})$ has large support with respect to the $L_1$ topology, and this is sufficient for posterior consistency. \citet{deyoreokottas} and \citet{Norets} show that ordinal regression models induced from mixtures of multivariate normals, similar to the model we use for $P(\mathbf{Y}^{(A)}\mid\mathbf{X})$, possess the Kullback-Leibler (KL) large support property. 
However, this does not imply that such properties are
  inherited by the joint model. Thus, the support and consistency properties of focused clustering models are  
topics for future research.  We note that large support and asymptotic results on posterior
  consistency for multivariate, mixed-scale distributions have been established by \citet{canale} and
  \citet{Norets}, but generally are few in comparison to the
  results available for continuous densities.  For MM-FC, one possible approach could involve extending the $L_1$ 
  support result for $P(\mathbf{X})$ to KL support, and using results of \citet{Norets} to obtain KL support for 
  $P(\mathbf{Y}^{(A)}\mid\mathbf{X})$.  The
  chain rule for relative entropy could then be applied to establish KL support for the
  joint model \citep{deyoreokottas}.  Regardless, we envision MM-FC to
  be most useful for modest-sized data, when focused clustering is most
  likely to be needed.  In such cases, it is crucial to check the quality
  of the model fit for the data at hand.

\bibliographystyle{ba}
\bibliography{biblioFS}

\begin{thebibliography}{46}
\newcommand{\enquote}[1]{``#1''}
\expandafter\ifx\csname natexlab\endcsname\relax\def\natexlab#1{#1}\fi
\expandafter\ifx\csname url\endcsname\relax
  \def\url#1{{\tt #1}}\fi
\expandafter\ifx\csname urlprefix\endcsname\relax\def\urlprefix{URL }\fi
\ifx\endbibitem\undefined \let\endbibitem\relax\fi

\bibitem[{Abayomi et~al.(2008)Abayomi, Gelman, and Levy}]{abayomi}
Abayomi, K., Gelman, A., and Levy, M. (2008).
\newblock \enquote{Diagnostics for multivariate imputations.}
\newblock {\em Journal of the Royal Statistical Society: Series C (Applied
  Statistics)\/}, 57(3): 273--291.
\endbibitem

\bibitem[{Albert and Chib(1993)}]{albert}
Albert, J. and Chib, S. (1993).
\newblock \enquote{Bayesian analysis of binary and polychotomous response
  data.}
\newblock {\em Journal of the American Statistical Association\/}, 88:
  669--679.
\endbibitem

\bibitem[{Banerjee et~al.(2013)Banerjee, Murray, and Dunson}]{banerjee}
Banerjee, A., Murray, J., and Dunson, D. (2013).
\newblock \enquote{Bayesian learning of joint distributions of objects.}
\newblock In {\em Proceedings of the 16th International Conference on
  Artificial Intelligence and Statistics\/}.
\endbibitem

\bibitem[{Bao and Hanson(2015)}]{bao}
Bao, J. and Hanson, T. (2015).
\newblock \enquote{Bayesian nonparametric multivariate ordinal regression.}
\newblock {\em Canadian Journal of Statistics\/}, 43: 337--357.
\endbibitem

\bibitem[{Bartles(1999)}]{bartels1999panel}
Bartles, L.~M. (1999).
\newblock \enquote{Panel effects in the American national election studies.}
\newblock {\em Political Analysis\/}, 8: 1--20.
\endbibitem

\bibitem[{Berinsky(2004)}]{berinsky2004silent}
Berinsky, A.~J. (2004).
\newblock {\em Silent Voices: Public Opinion and Political Participation in
  America\/}.
\newblock Princeton University Press.
\endbibitem

\bibitem[{Bishop et~al.(1975)Bishop, Fienberg, and Holland}]{bishop}
Bishop, Y., Fienberg, S., and Holland, P. (1975).
\newblock {\em Discrete Multivariate Analysis: Theory and practice\/}.
\newblock Cambridge, MA: M.I.T. Press.
\endbibitem

\bibitem[{Boes and Winkelmann(2006)}]{boes}
Boes, S. and Winkelmann, R. (2006).
\newblock {\em Ordered Response Models\/}, 167--181.
\newblock Springer Berlin Heidelberg.
\endbibitem

\bibitem[{B{\"o}hning et~al.(2007)B{\"o}hning, Seidel, Alfo, Garel, Patilea,
  Walther, Zio, Guarnzera, and Luzi}]{bohning}
B{\"o}hning, D., Seidel, W., Alfo, M., Garel, B., Patilea, V., Walther, G.,
  Zio, M.~D., Guarnzera, U., and Luzi, O. (2007).
\newblock \enquote{Imputation through finite Gaussian mixture models.}
\newblock {\em Computational Statistics and Data Analysis\/}, 51: 5305--5316.
\endbibitem

\bibitem[{Canale and Dunson(2015)}]{canale}
Canale, A. and Dunson, D. (2015).
\newblock \enquote{Bayesian multivariate mixed-scale density estimation.}
\newblock {\em Statistics and its Interface\/}, 8: 195--201.
\endbibitem

\bibitem[{Chib and Greenberg(1998)}]{chibgreen}
Chib, S. and Greenberg, E. (1998).
\newblock \enquote{Analysis of multivariate probit models.}
\newblock {\em Biometrika\/}, 85: 347--361.
\endbibitem

\bibitem[{DeYoreo and Kottas(2014)}]{deyoreokottas}
DeYoreo, M. and Kottas, A. (2014).
\newblock \enquote{Bayesian nonparametric modeling for multivariate ordinal
  regression.}
\newblock {\em arXiv:1408.1027\/}, stat.ME.
\endbibitem

\bibitem[{DeYoreo and Kottas(2015)}]{deyoreo:binary}
--- (2015).
\newblock \enquote{A fully nonparametric modeling approach to binary
  regression.}
\newblock {\em Bayesian Analysis\/}, 10: 821--847.
\endbibitem

\bibitem[{Dunson and Bhattacharya(2010)}]{dunson:bhat}
Dunson, D. and Bhattacharya, A. (2010).
\newblock \enquote{Nonparametric Bayes regression and classication through
  mixtures of product kernels.}
\newblock In Bernardo, J.~M., Bayarri, M.~J., Berger, J.~O., Dawid, A.~P.,
  Heckerman, D., Smith, A. F.~M., and West, M. (eds.), {\em Bayesian Statistics
  9, Proceedings of Ninth Valencia International Conference on Bayesian
  Statistics\/}, 145--164.
\endbibitem

\bibitem[{Dunson and Xing(2009)}]{dunsonxing}
Dunson, D. and Xing, C. (2009).
\newblock \enquote{Nonparametric Bayes modeling of multivariate categorical
  data.}
\newblock {\em Journal of the American Statistical Association\/}, 104:
  1042--1051.
\endbibitem

\bibitem[{Elliott and Stettler(2007)}]{elliott}
Elliott, M. and Stettler, N. (2007).
\newblock \enquote{Using a mixture model for multiple imputation in the
  presence of outliers: the Healthy for Life project.}
\newblock {\em Journal of the Royal Statistical Society: Series C\/}, 56:
  63--78.
\endbibitem

\bibitem[{Gelman et~al.(2005)Gelman, Van{ }Mechelen, Verbeke, and
  Meulders}]{gelman-model-checking}
Gelman, A., Van{ }Mechelen, I., Verbeke, G., and Meulders, H. (2005).
\newblock \enquote{Multiple imputation for model checking: completed-data plots
  with missing and latent data.}
\newblock {\em Biometrics\/}, 61: 74--85.
\endbibitem

\bibitem[{Ghahramani and Hinton(1997)}]{ghahramani}
Ghahramani, Z. and Hinton, G. (1997).
\newblock \enquote{The EM algorithm for mixtures of factor analyzers.}
\newblock Technical report, University of Toronto.
\endbibitem

\bibitem[{Gorur and Rasmussen(2009)}]{gorur}
Gorur, D. and Rasmussen, C. (2009).
\newblock \enquote{Nonparametric mixtures of factor analyzers.}
\newblock {\em Sigma Processing and Communications Applications Conference\/},
  708--711.
\endbibitem

\bibitem[{Hannah et~al.(2011)Hannah, Blei, and Powell}]{hannah}
Hannah, L., Blei, D., and Powell, W. (2011).
\newblock \enquote{Dirichlet process mixtures of generalized linear models.}
\newblock {\em Journal of Machine Learning Research\/}, 1: 1--33.
\endbibitem

\bibitem[{He and Zaslavsky(2012)}]{zaslavsky}
He, Y. and Zaslavsky, A. (2012).
\newblock \enquote{Diagnosing imputation models by applying target analyses to
  posterior replicates of completed data.}
\newblock {\em Statistics in Medicine\/}, 31: 1--18.
\endbibitem

\bibitem[{Honaker et~al.(2011)Honaker, King, and Blackwell}]{Honaker}
Honaker, J., King, G., and Blackwell, M. (2011).
\newblock \enquote{Amelia II: A program for missing data.}
\newblock {\em Journal of Statistical Software\/}, 45(7): 1--47.
\endbibitem

\bibitem[{Ibrahim et~al.(1999)Ibrahim, Lipsitz, and Chen}]{ibrahim:1999}
Ibrahim, J., Lipsitz, S., and Chen, M. (1999).
\newblock \enquote{Missing covariates in generalized linear models when the
  missing data mechanism is non-ignorable.}
\newblock {\em Journal of the Royal Statistical Society, Series B\/}, 61:
  173--190.
\endbibitem

\bibitem[{Kim et~al.(2015)Kim, Cox, Karr, Reiter, and Wang}]{kim:JASA}
Kim, H.~J., Cox, L., Karr, A., Reiter, J., and Wang, Q. (2015).
\newblock \enquote{Simultaneous edit-imputation for continuous microdata.}
\newblock {\em Journal of the American Statistical Association\/}, 110:
  987--999.
\endbibitem

\bibitem[{Kim et~al.(2014)Kim, Reiter, Wang, Cox, and Karr}]{kim}
Kim, H.~J., Reiter, J.~P., Wang, Q., Cox, L., and Karr, A. (2014).
\newblock \enquote{Multiple imputation of missing or faulty values under linear
  constraints.}
\newblock {\em Journal of Business and Economic Statistics\/}, 32: 375--386.
\endbibitem

\bibitem[{Kottas et~al.(2005)Kottas, M{\"u}ller, and Quintana}]{kottas}
Kottas, A., M{\"u}ller, P., and Quintana, F. (2005).
\newblock \enquote{Nonparametric Bayesian modelling for multivariate ordinal
  data.}
\newblock {\em Journal of Computational and Graphical Statistics\/}, 14:
  610--625.
\endbibitem

\bibitem[{Lipsitz and Ibrahim(1996)}]{lipsitz}
Lipsitz, S. and Ibrahim, J. (1996).
\newblock \enquote{A conditional model for incomplete covariates in parametric
  regression models.}
\newblock {\em Biometrika\/}, 83: 916--922.
\endbibitem

\bibitem[{Little and Rubin(2002)}]{little:rubin:2002}
Little, R. and Rubin, D. (2002).
\newblock {\em Statistical Analysis with Missing Data\/}.
\newblock New York: Wiley.
\endbibitem

\bibitem[{Manrique-Vallier and Reiter(2014)}]{manrique}
Manrique-Vallier, D. and Reiter, J. (2014).
\newblock \enquote{Bayesian multiple imputation for large-scale categorical
  data with structural zeros.}
\newblock {\em Survey Methodology\/}, 40: 125--134.
\endbibitem

\bibitem[{McParland et~al.(2014)McParland, Gormley, McCormick, Clark, Whiteson,
  and Collinson}]{mcparland}
McParland, D., Gormley, I., McCormick, T., Clark, S., Whiteson, K., and
  Collinson, M. (2014).
\newblock \enquote{Clustering South African households based on their asset
  status using latent variable models.}
\newblock {\em Annals of Applied Statistics\/}, 8: 747--776.
\endbibitem

\bibitem[{M{\"u}ller and Mitra(2013)}]{muller:mitra}
M{\"u}ller, P. and Mitra, R. (2013).
\newblock \enquote{Bayesian nonparametric inference: Why and how?}
\newblock {\em Bayesian Analysis\/}, 8: 269--302.
\endbibitem

\bibitem[{Murray and Reiter(2016)}]{murray}
Murray, J. and Reiter, J. (2016).
\newblock \enquote{Multiple imputation of missing categorical and continuous
  values via Bayesian mixture models with local dependence.}
\newblock {\em Journal of the American Statistical Association\/}, To appear.
\endbibitem

\bibitem[{Norets and Pelenis(2012)}]{Norets}
Norets, A. and Pelenis, J. (2012).
\newblock \enquote{Bayesian modeling of joint and conditional distributions.}
\newblock {\em Journal of Econometrics\/}, 168: 332--346.
\endbibitem

\bibitem[{Pasek et~al.(2009)Pasek, Tahk, Lelkes, Krosnick, Payne, Akhtar, and
  Tompson}]{pasek2009determinants}
Pasek, J., Tahk, A., Lelkes, Y., Krosnick, J.~A., Payne, B.~K., Akhtar, O., and
  Tompson, T. (2009).
\newblock \enquote{Determinants of turnout and candidate choice in the 2008
  U.S. presidential election illuminating the impact of racial prejudice and
  other considerations.}
\newblock {\em Public Opinion Quarterly\/}, 73(5): 943--994.
\endbibitem

\bibitem[{Peress(2010)}]{peress2010correcting}
Peress, M. (2010).
\newblock \enquote{Correcting for survey nonresponse using variable response
  propensity.}
\newblock {\em Journal of the American Statistical Association\/}, 105:
  1418--1430.
\endbibitem

\bibitem[{Petralia et~al.(2012)Petralia, Rao, and Dunson}]{Petralia}
Petralia, F., Rao, V., and Dunson, D. (2012).
\newblock \enquote{Repulsive mixtures.}
\newblock {\em Advances in Neural Information Processing Systems\/}, 25.
\endbibitem

\bibitem[{Raghunathan et~al.(2001)Raghunathan, Lepkowski, Van~Hoewyk, and
  Solenberger}]{raghunathan2001multivariate}
Raghunathan, T.~E., Lepkowski, J.~M., Van~Hoewyk, J., and Solenberger, P.
  (2001).
\newblock \enquote{A multivariate technique for multiply imputing missing
  values using a sequence of regression models.}
\newblock {\em Survey Methodology\/}, 27(1): 85--96.
\endbibitem

\bibitem[{Reiter and Raghunathan(2007)}]{reiterrag}
Reiter, J.~P. and Raghunathan, T.~E. (2007).
\newblock \enquote{The multiple adaptations of multiple imputation.}
\newblock {\em Journal of the American Statistical Association\/}, 102:
  1462--1471.
\endbibitem

\bibitem[{Rubin(1987)}]{rubin}
Rubin, D. (1987).
\newblock {\em Multiple Imputation for Nonresponse in Surveys\/}.
\newblock New York: John Wiley and Sons.
\endbibitem

\bibitem[{Rubin(1996)}]{rubin1996}
--- (1996).
\newblock \enquote{Multiple imputation after 18+ years.}
\newblock {\em Journal of the American Statistical Association\/}, 91:
  473--489.
\endbibitem

\bibitem[{Si and Reiter(2013)}]{si}
Si, Y. and Reiter, J. (2013).
\newblock \enquote{Nonparametric Bayesian multiple imputation for incomplete
  categorical variables in large-scale assessment surveys.}
\newblock {\em Journal of Educational and Behavioral Statistics\/}, 38:
  499--521.
\endbibitem

\bibitem[{Treier and Hillygus(2009)}]{treier2009}
Treier, S. and Hillygus, D. (2009).
\newblock \enquote{The nature of political ideology in the contemporary
  electorate.}
\newblock {\em Public Opinion Quarterly\/}, 73: 679--703.
\endbibitem

\bibitem[{van{ }Buuren and Groothuis-Oudshoorn(2011)}]{vanbuuren2011}
van{ }Buuren, S. and Groothuis-Oudshoorn, K. (2011).
\newblock \enquote{Mice: Multivariate imputation by chained equations.}
\newblock {\em Journal of Statistical Software\/}, 45(3): 1--67.
\endbibitem

\bibitem[{Vermunt et~al.(2008)Vermunt, Ginkel, der Ark, and Sijtsma}]{vermunt}
Vermunt, J., Ginkel, J., der Ark, L., and Sijtsma, K. (2008).
\newblock \enquote{Multiple imputation of incomplete categorical data using
  latent class analysis.}
\newblock {\em Sociological Methodology\/}, 38: 369--397.
\endbibitem

\bibitem[{Wade et~al.(2014{\natexlab{a}})Wade, Dunson, Perone, and
  Trippa}]{wadedunson}
Wade, S., Dunson, D., Perone, S., and Trippa, L. (2014{\natexlab{a}}).
\newblock \enquote{Improving prediction from Dirichlet process mixtures via
  enrichment.}
\newblock {\em Journal of Machine Learning Research\/}, 15: 1041--1071.
\endbibitem

\bibitem[{Wade et~al.(2014{\natexlab{b}})Wade, Walker, and
  Petrone}]{wadewalker}
Wade, S., Walker, S.~G., and Petrone, S. (2014{\natexlab{b}}).
\newblock \enquote{A predictive study of Dirichlet process mixture models for
  curve fitting.}
\newblock {\em Scandinavian Journal of Statistics\/}, 41: 580--605.
\endbibitem

\end{thebibliography}

\begin{acknowledgement}
We thank two referees, the Associate Editor, and the Editor for comments and suggestions that greatly improved this manuscript. 
\end{acknowledgement}

\end{document}